\documentclass[usenatbib]{mn2e}
\usepackage{graphicx}
\usepackage{epstopdf}
\usepackage{natbib}
\usepackage{verbatim}
\usepackage{amssymb,amsmath}
\usepackage{ulem}
\usepackage{color}
\usepackage{pifont}
\usepackage[bookmarks,bookmarksnumbered,colorlinks=true, citecolor=blue, linkcolor=black]{hyperref}

\pdfoutput=1

\def \ltsim {\lesssim}

\renewcommand{\thefootnote}{\fnsymbol{footnote}}

\newcommand\aj{AJ}
\newcommand\araa{ARA\&A}
\newcommand\apj{ApJ}
\newcommand\apjl{ApJL}
\newcommand\apjs{ApJS}
\newcommand\aap{A\&A}
\newcommand\mnras{MNRAS}

\newcommand\altaffilmark[1]{$^{#1}$}
\newcommand\altaffiltext[1]{$^{#1}$}
\newcommand{\etal}{et al.}

\title[Unstable feedback in galactic nuclei]{An instability of feedback regulated star formation in galactic nuclei}

\vspace{-0.2cm}
\author[Torrey \etal]{
\parbox[t]{\textwidth}{ 
Paul~Torrey\thanks{E-mail: ptorrey@mit.edu}\altaffilmark{1,2},
Philip F.~Hopkins\altaffilmark{2}, 
Claude-Andr{\'e} Faucher-Gigu{\`e}re\altaffilmark{3}, 
 \\ Mark Vogelsberger\altaffilmark{1},
Eliot Quataert\altaffilmark{4},
Du\v{s}an Kere\v{s}\altaffilmark{5},
\& Norman Murray\altaffilmark{6,7}
} 
\vspace*{6pt} \\
\altaffiltext{1}{MIT Kavli Institute for Astrophysics \& Space Research, Cambridge, MA, 02139, USA} \\
\altaffiltext{2}{TAPIR, Mailcode 350-17, California Institute of Technology, Pasadena, CA 91125, USA} \\
\altaffiltext{3}{Department of Physics and Astronomy and CIERA, Northwestern University, 2145 Sheridan Road, Evanston, IL 60208, USA} \\ 
\altaffiltext{4}{Department of Astronomy and Theoretical Astrophysics Center, University of California Berkeley, Berkeley, CA 94720, USA}\\
\altaffiltext{5}{Department of Physics, University of California at San Diego, 9500 Gilman Drive, La Jolla, CA 92093, USA}\\
\altaffiltext{6}{Canadian Institute for Theoretical Astrophysics, 60 St. George Street, University of Toronto, ON M5S 3H8, Canada}\\
\altaffiltext{7}{Canada Research Chair in Astrophysics}
\vspace{-0.5cm}
}

\begin{document}

\maketitle

\begin{abstract}
We examine the stability of feedback-regulated star formation (SF) in galactic nuclei and contrast it to SF in extended discs. 
In galactic nuclei the dynamical time becomes shorter than the time over which feedback from young stars evolves. 
We argue analytically that the balance between stellar feedback and gravity is unstable in this regime.
We study this using numerical simulations with pc-scale resolution and explicit stellar feedback taken from stellar evolution models. 
The nuclear gas mass, young stellar mass, and SFR within the central $\sim$100 pc (the short-timescale regime) never reach steady-state, but instead go through dramatic, oscillatory cycles. 
Stars form until a critical surface density of young stars is present (such that feedback overwhelms gravity), at which point they begin to expel gas from the nucleus. 
Since the dynamical times are shorter than the stellar evolution times,
the stars do not die as the gas is expelled, but continue to push, triggering a runaway quenching of star formation in the nucleus. 
However the expelled gas is largely not unbound from the galaxy, but goes into a galactic fountain which re-fills the nuclear region after the massive stars from the previous burst cycle have died off ($\sim$50 Myr timescale). 
On large scales ($>$1 kpc), the galaxy-scale gas content and SFR is more stable.
We examine the consequences of this episodic nuclear star formation for the Kennicutt-Schmidt (KS) relation: while a tight KS relation exists on $\sim$1 kpc scales in good agreement with observations, the scatter increases dramatically in smaller apertures centered on galactic nuclei.
\end{abstract}


\renewcommand{\thefootnote}{\fnsymbol{footnote}}

\section{Introduction}
Observations of star formation in our Galaxy and in extragalactic systems show that stars are formed inefficiently.
A common approach for parameterizing the star formation rate (SFR) efficiency is to express
$\dot \rho_*  = \epsilon \rho_{{\rm gas}}/t_{{\rm ff}}$
where $\dot \rho_*$ is the star formation rate per unit volume, $\rho_{{\rm gas}}$ is the volume density of gas, $t_{{\rm ff}}$ is the local gas free fall time, and $\epsilon$ is the star formation rate efficiency (i.e. the fraction of gas that turns into stars per galactic free fall time).
Since the volumetric gas and star formation rate densities are not easily observable, this inefficiency is observationally probed through the Kennicutt-Schmidt~\citep[KS;][]{Schmidt, Kennicutt} relation.
The KS relation considers the current star formation rate surface density as a function of the available fuel for star formation measured through the total (or molecular) gas surface density.
Although there is still debate about the detailed slope and normalization of the KS relation, it is generally agreed that -- when measured/averaged over large areas within galaxies -- only a few percent of gas is able to be converted into stars per free fall time~\citep[e.g.,][]{Zuckerman1974, Leroy2008, Bigiel2008}.
Understanding and explaining this inefficiency forms a central component of star formation and galaxy formation research.

Two main classes of models have been presented to explain the low efficiency of star formation and the form of the KS law.
In one class it is argued that the small-scale properties of supersonic ISM turbulence are able to disrupt cloud collapse, giving rise to low SFR efficiencies~\citep[e.g.,][]{Krumholz2005, Federrath2012}.
These models relate the low efficiency of star formation to the statistical properties of the density fluctuations in supersonic turbulence.
However, since the rate-limiting step for gravitational collapse is the formation of dense gas clouds in the first place, this is an incomplete explanation of the galaxy-scale KS law.

Another approach has been to develop analytic global ``equilibrium" models that explain the low efficiency of star formation and form of the KS law~\citep{Silk1997, Thompson2005, Ostriker2011, FaucherGiguere2013}.
The common component of the equilibrium models is that the interstellar medium (ISM) remains in a quasi stable state by balancing feedback from young stellar populations with the global pressure of the self gravity of the disk~\citep{Silk1997, Hopkins2011a}.
While thermal pressure or radiation pressure contribute to the disk support, play a role in disrupting the giant molecular cloud (GMC) complexes where stars are formed~\citep[e.g.,][]{Murray2010}, and could be a source of turbulence, turbulent pressure is thought to provide the dominant source of support against the vertical collapse in gas-rich galaxies~\citep{Ostriker2011, FaucherGiguere2013}, except possibly in the most optically thick inner parts~\citep{Thompson2005}.
Equilibrium star formation models are able to derive star formation efficiencies by asserting that the turbulent energy injection rate from young stellar populations balances the turbulent energy dissipation rate.
The derived star formation efficiencies scale with measurable properties of galactic disks and are consistent with observed star formation efficiencies~\citep{FaucherGiguere2013}.

These models implicitly assume that 
(1) gas in the local disk responds to feedback, 
(2) the strength of feedback scales with the star formation {\it rate}, and 
(3) the star formation rate -- and therefore feedback -- responds in turn to the gas properties. 
If feedback is ``too strong," gas will be expelled or pushed out to such low densities that it cannot form new stars, so the feedback ``supply" will cease. 
If feedback is ``too weak," gas collapses rapidly and forms more stars, injecting more feedback until collapse is halted. 

This feedback cycle can only hold if the gas adjusts efficiently to the presence of feedback, {\it and the feedback energy/momentum injection rate efficiently adjusts} to changes in the gas density via the star formation rate.
As long as the feedback is sufficiently strong and efficiently coupled to the gas, the gas will adjust to the presence of feedback.
However, there are physical regimes where feedback evolves on timescales that are (much) longer than the local dynamical time.
In these regimes, the feedback energy/momentum injection rate will not respond to changes in the gas density, potentially breaking the feedback loop invoked in these star formation models.

For stellar feedback, the primary sources of direct momentum and energy injection into the local environment are radiation pressure, stellar winds, photoionization, and supernovae~\citep{Murray2010, HopkinsKSlaw}, and cosmic rays~\citep[e.g.,][]{Jubelgas2008, Uhlig2012, Booth2013, Salem2014}.   Magnetic fields may also act to suppress star formation regulation mechanisms~\citep{Dolag1999,Wang2009, Pakmor2013, Marinacci2015}.
The relative importance of these mechanisms depends on the physical conditions of the gas.
Radiation pressure~\citep{Murray2010} and photoionization~\citep{Whitworth1979, Krumholz2006, Walch2012, Sales2014} are often identified as being responsible for disrupting the dense star forming gas,
but supernovae probably dominate the turbulent momentum injection that balances the disk vertical collapse under most conditions~\citep{Ostriker2011}.

\begin{figure}
\centerline{\vbox{\hbox{\includegraphics[width=0.475\textwidth]{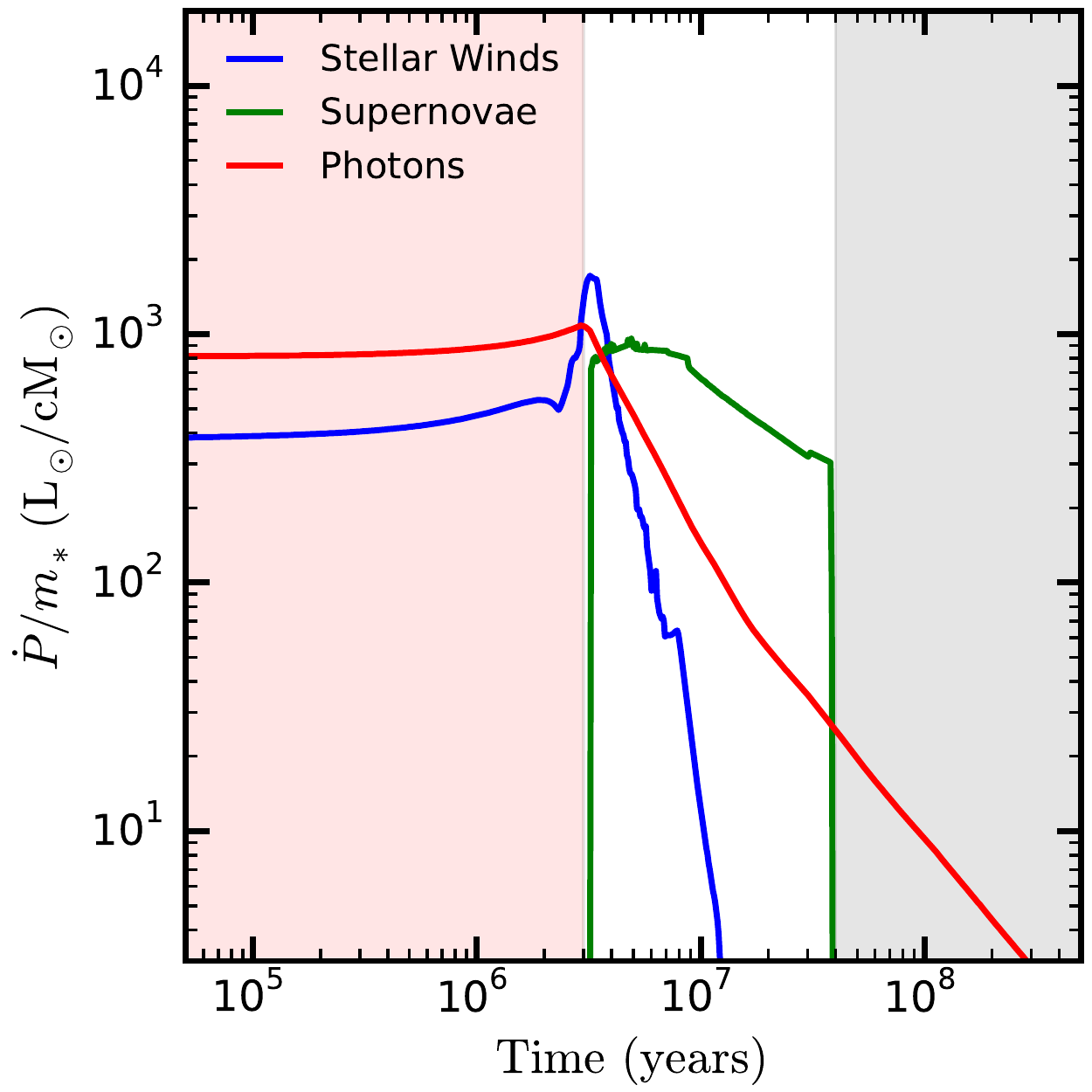}}}}
\caption{The direct momentum input taken from the {\small STARBURST99} stellar evolution models using a Kroupa IMF. 
The first $\sim3 \times 10^6$ years are characterized by nearly constant feedback levels dominated by radiation pressure from young stars.
The direct momentum injections levels begin to evolve only after massive stars begin moving off the main sequence which leads to a continual decrease in the direct radiation pressure momentum injection and an increase in the supernova momentum injection.  
We note that the momentum injection values shown are direct momentum injection values and do not account 
for any gains owing to trapping of photons (i.e. multiple scattering events), gains from pressure confinement of hot gas (the ``b" factor, as defined in Equation~\ref{eqn:new_pstar}), or losses owing to inefficient coupling.
We note that the boost to the direct momentum injection is likely modest for radiation pressure owing to radiation hydro instabilities, but can provide an order of magnitude increase in the supernova contribution to the total momentum injection.}
\label{fig:sb99}
\end{figure}

Figure~\ref{fig:sb99} shows the direct momentum injection budget associated with a single age stellar population as a function of time since birth taken directly from {\small STARBURST99}~\citep{SB99_3}.
Radiation pressure dominates the early stellar feedback budget until the death of the most massive stars ($\sim$3 Myr) and drops to $\sim$10\% of its initial value by $\sim$10 Myr.
Supernovae dominate the direct momentum injection budget starting at $\sim$3 Myr and continue to dominate until the time after which the least massive stars that die in supernovae move off the main sequence ($\sim$40 Myr)\footnote{For a Kroupa IMF with $M=8M_\odot$ assumed to be the lower mass limit for supernova progenitors.}.
From Figure~\ref{fig:sb99}, the total momentum injection rate remains nearly constant until $\sim$3 Myr, and remains variable but order-of-magnitude constant until $\sim$40 Myr.
The constant early momentum injection rates allow us to define an interesting regime:  regions of galaxies that possess dynamical times less than the timescale over which stellar feedback levels adjust.
In these regimes, feedback can not efficiently adjust to changes in the gas distribution and/or the current star formation rate.
In this paper we focus on the nuclear regions of galaxies where the small spatial scales and central black hole ensure consistently short dynamical times.

The understanding of star formation in galactic nuclei that we present in this paper will set the stage for follow up studies focused on the accretion, growth, and feedback associated with supermassive black holes.
Previous studies have considered the impact of stellar and black hole feedback in concert on the central $\sim$100 pc nuclear disk~\citep[e.g.,][]{Hopkins2015}.
However, those studies lacked the large-scale galactic disk that is important for understanding the long term behavior of supermassive black hole accretion and feedback.

The structure of this paper is as follows:
In Section~\ref{sec:AA} we review the basic components of equilibrium feedback models.
This allows us to concretely explore the assumptions that break down for the equilibrium model at short dynamical times.
We propose alterations to the equilibrium feedback models that hold within short dynamical time regions and conclude that star formation within short dynamical time regions is likely to be inherently bursty.
In Section~\ref{sec:Simulations} we describe the numerical simulations that we use to explore the properties of nuclear star formation.
In Section~\ref{sec:Results} we present the results of our numerical simulations, with a focus on the bursty nature of nuclear star formation and implications for observations of the KS relation.
In Section~\ref{sec:Discussion} we present a discussion of our results with implications for observations of nuclear stellar disks as well as the observed KS relation.
In Section~\ref{sec:Conclusions} we summarize and conclude.

\section{Analytic Arguments}
\label{sec:AA}
\subsection{The Equilibrium Model}
Analytic models have been developed to explain the efficiency of star formation and form of the KS law.  
In global equilibrium models the star formation efficiency is derived by asserting that turbulent energy injection from young stellar populations balances the turbulent energy dissipation rate.
One of the first models to do this was~\citet{Thompson2005}, where they examined radiation and mechanical feedback from young stellar populations with a focus on the central 100 pc of a starburst disk.
We adopt a similar setup here, which includes an isothermal spherical potential -- which could represent either a halo or bulge component.
The projected surface density follows a power law $\Sigma_{\rm tot} = \sigma^2 / \pi G r $
where $\sigma$ is the velocity dispersion of the potential.  
The mass profile is given by $M_{\rm tot} = 2 \sigma^2 r /G $, with an angular frequency of $\Omega = \sqrt{2} \sigma / r$ set to enforce radial centrifugal balance.
The specified gas fraction is $f_{\rm g} = M_{\rm g}/M_{ {\rm tot} } =  \Sigma_{\rm g}/\Sigma_{ {\rm tot} }$.
For such a model the disk mid plane pressure required to balance gravity can be approximated by
\begin{equation}
p \approx \rho_{\rm g} h^2 \Omega^2.
\label{eqn:mid_plane_pressure}
\end{equation}
where $h$ is the disk scale height and $\rho_{\rm g}$ is the gas density.
We then assume that supersonic turbulent pressure is the primary source of pressure support in the disk with turbulent gas velocity dispersion $c_t$, in which case the rate of (volumetric) turbulent energy dissipation is
\begin{equation}
\dot u  \approx \rho_{\rm g}  c_t^2 \Omega.
\label{eqn:udot_diss}
\end{equation}
Under the assumption that the system approaches an equilibrium, this energy dissipation rate needs to be balanced by feedback from young stars.
There are several possible ways to write down the feedback injection rate.  
Here, we express the energy injection rate from young stars as 
\begin{equation}
\dot u \approx \frac{p_* c_t}{h} = \frac{ f_p P_*/m_* \dot \Sigma_* c_t }{ 4 h}.
\end{equation}
The second equality follows from expressing the turbulent pressure produced by stellar feedback as
\begin{equation}
p_* = f_p \frac{P_*}{4 m_*} \dot \Sigma_*
\label{eqn:sfr_pressure}
\end{equation}
where $P_*/m_*$ is the radial momentum injected per unit stellar mass formed~\citep{Ostriker2011, FaucherGiguere2013} and $f_p$ parameterizes the efficiency with which the radial momentum drives turbulent energy injection, but is of likely of order unity~\citep{Ostriker2011}. 
Adopting the above expression of the young star pressure/turbulent energy injection rate does not require any assumptions about the origin of the momentum input, and can accommodate multiple momentum input sources (e.g., radiation pressure, supernovae, stellar winds, etc.).
For SNe, typical values for $P_*/m_*$ are of order $\sim 3000$ km s${}^{-1}$~\citep[see][for additional details]{Ostriker2011}.

Enforcing pressure balance between the self gravity of the disk and turbulent pressure (i.e. between Equations~\ref{eqn:mid_plane_pressure} and~\ref{eqn:sfr_pressure}) we find the feedback self-regulated equilibrium condition 
\begin{equation}
\rho_{\rm g}  c_t ^2  \approx  \frac{ f_p P_*/m_*\dot \Sigma_*  }{ 4  }.
\end{equation}
Using $c_t = h \Omega$ and $ \Sigma_{\rm g}=2 h \rho_{\rm g}$, this can be rearranged in terms of the equilibrium star formation rate surface density
\begin{equation}
\dot \Sigma_* \approx  \eta \Sigma_{\rm g} \Omega = \Sigma_{\rm g} \Omega \frac{ 2 c_t     }{  f_p P_*/m_*  }
\end{equation}
where $\eta = 2 c_t  /   f_p (P_*/m_*)  $ is an efficiency that describes the fraction of gas that can convert into stars in a dynamical time.
This expression shows that the galactic star formation efficiency is directly related to the gas turbulent velocity dispersion in the disk.

\begin{figure*}
\centerline{\vbox{\hbox{
\includegraphics[width=0.333\textwidth]{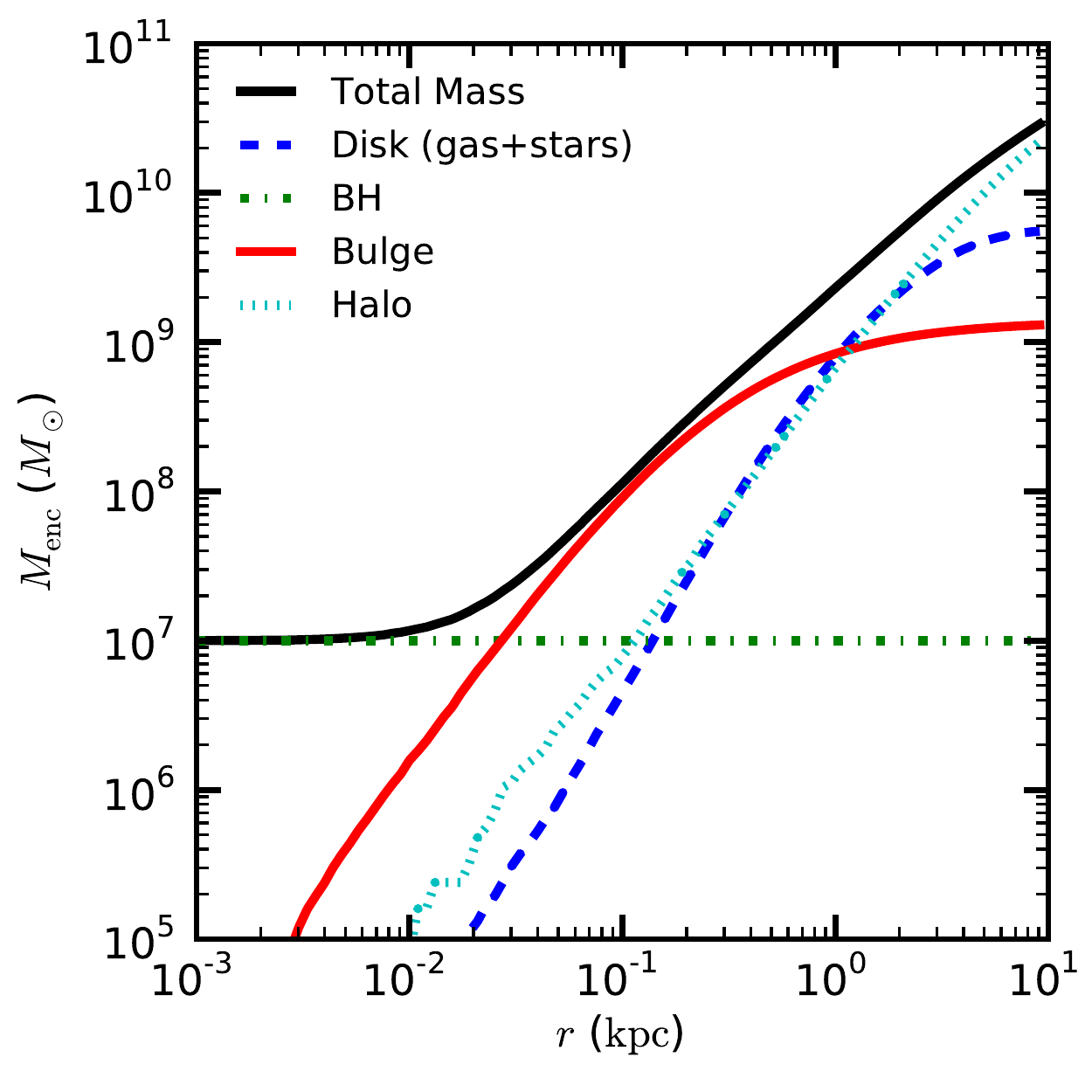}
\includegraphics[width=0.333\textwidth]{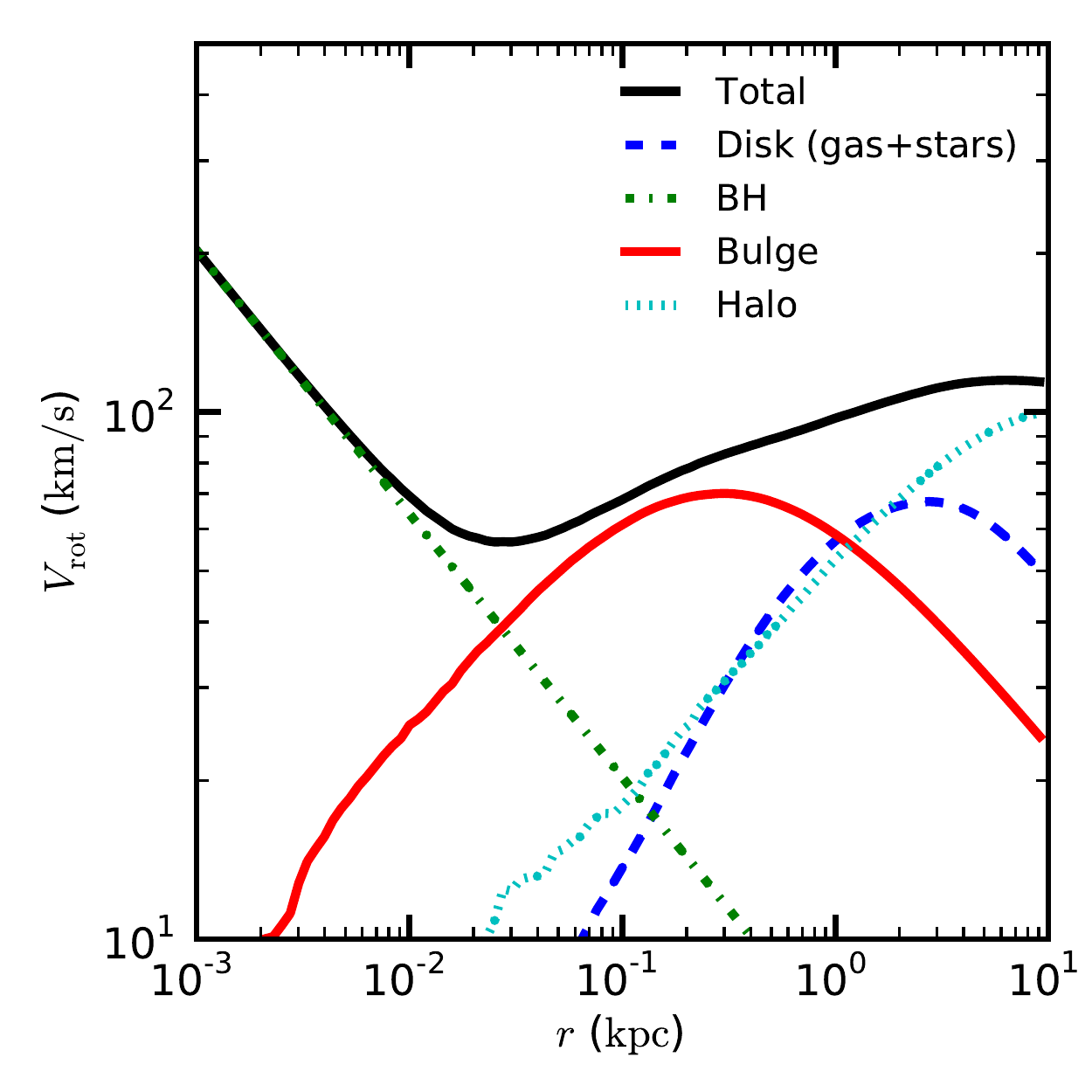}
\includegraphics[width=0.333\textwidth]{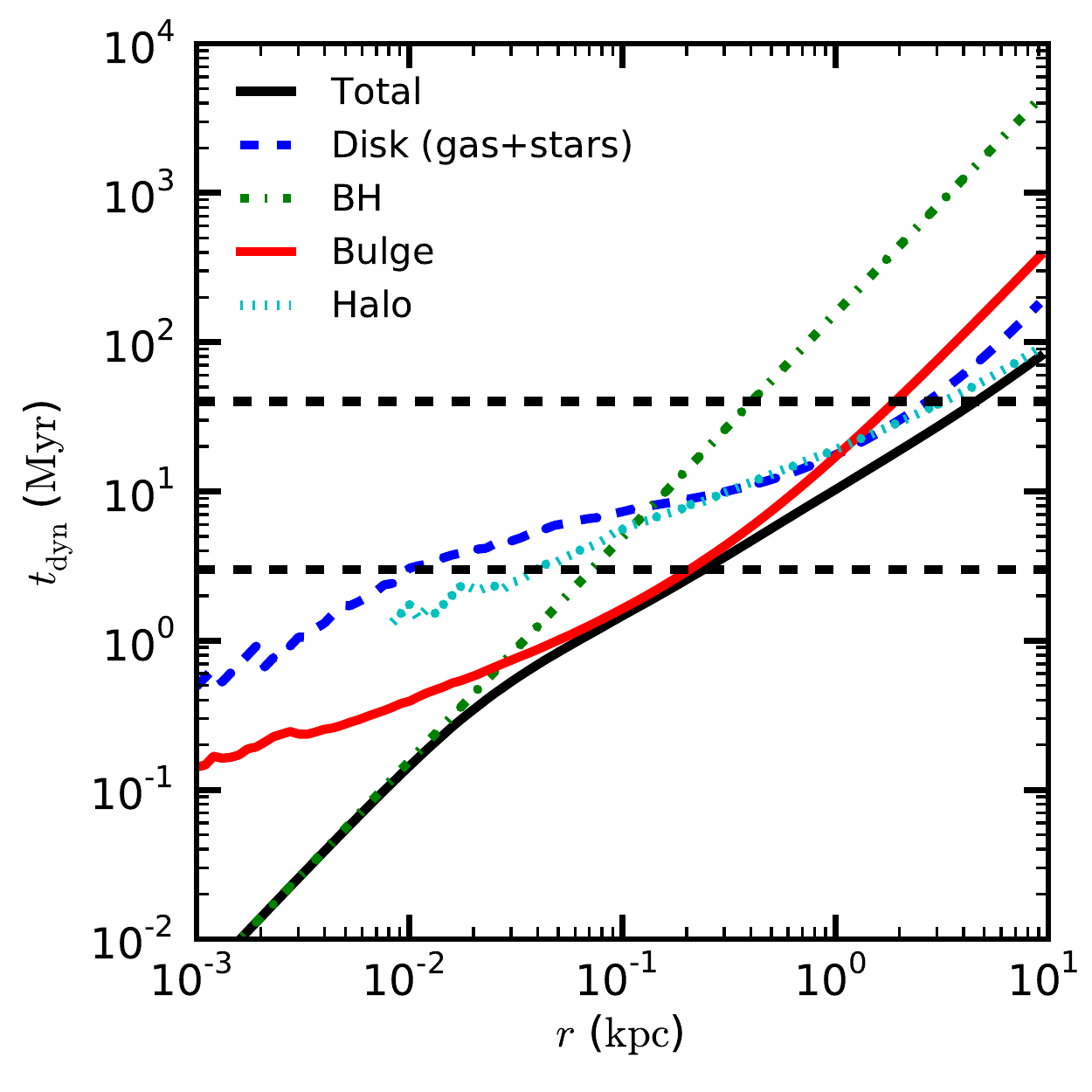}
}}}
\caption{The mass enclosed (left), rotational velocity (center), and dynamical time (right) are shown as a function of radius for the D1 (low mass) initial disk.  The black hole's dynamical relevance at scales $r\ltsim$10-30 parsecs can be identified from any of the plots.   The horizontal lines in the dynamical time plot indicate 3 and 40 Myr.  Dynamical times are calculated using the enclosed mass profiles to determine the circular velocities.}
\label{fig:ICprofiles}
\end{figure*}

\subsection{Where the equilibrium models break down}
The equilibrium model as described above breaks down in two regimes.

\subsubsection{Equilibrium Breakdown from Short Dynamical Times}
The equilibrium model breaks down when feedback is unable to modulate its strength on sufficiently short timescales (i.e. the gas dynamical timescale).
Feedback levels are set by young stellar populations, which evolve based on the rate at which massive stars evolve.
Here, we are specifically concerned with the timescale over which radiation pressure and energy/momentum injection from supernovae decline as an initially young stellar population ages.
Radiation pressure begins to drop when a stellar population is just $\sim 3$ Myr old while supernova inject significant momentum until $\sim 40$ Myr (Figure~\ref{fig:sb99}).
Both of these feedback mechanisms have fallen to a small fraction ($\sim0.01$) of their initial value by the time a stellar population is $\sim100$ Myr old.
In regimes where the dynamical time is much shorter than the stellar evolution time (i.e. $\tau_{{\rm SE}} \sim 10^6-10^7$ years; see Figure~\ref{fig:sb99})\footnote{In what follows, our analysis does not require that we identify a specific value of $\tau_{{\rm SE}}$.}, 
feedback will not react sufficiently quickly to changes in the current gas density or star formation rate, but rather remain strong for a stellar evolution timescale.

For the one zone model used in this section, the dynamical time is
\begin{equation}
\tau_{{\rm dyn}} = \frac{\sqrt{2} \pi r }{\sigma} \sim 10^6 {\rm yr}  \left( \frac{r}{100 {\rm pc} }  \right) \left(  \frac{\sigma}{200 {\rm km/s} }   \right)^{-1}.
\end{equation}
Comparing this with a stellar evolution timescale $\tau_{{\rm SE}}$ we find that the dynamical time will be much shorter than the stellar evolution time ($\tau_{{\rm dyn}} \ll \tau_{{\rm SE}}$) when
\begin{equation}
r   \ll 1  {\rm kpc}  \left(  \frac{\tau_{{\rm SE}}}{10 {\rm Myr}}    \right) \left(  \frac{\sigma}{200 {\rm km/s} }   \right).
\end{equation}
When this condition is met, massive stars contributing significantly to SNe and radiation pressure live longer than the gas dynamical time and thus feedback from young stars will persist for several dynamical times.
For reasonable $\tau_{{\rm SE}}$ values (e.g. $\sim10$ Myr) the dynamical time can be much shorter than $\tau_{{\rm SE}}$ out to a significant fraction of the central $1$ kpc.
If the feedback from young stars exceeds gravity within this region, 
then the system becomes unstable/unbound and remains in that state for several dynamical times until the young stars responsible for feedback move off the main sequence.

To explore this regime, we define the convenient quantity $a_*(t)=\dot P / m_*$, which has units of acceleration. 
This quantity is plotted in Figure 1 as a function of time. 
We can express the turbulent pressure provided by young stars as
\begin{equation}
p_{*,y} = \Sigma_{*,y} a_*(t) \cdot b f_p/4,
\end{equation}
where  $\Sigma_{*,y}$ is the surface density of ``young stars" which dominate the stellar feedback budget.  
Throughout, we adopt a definition of young stars as being those stars with ages less than 40 Myrs.  
The quantity $b$ accounts for the increase in the momentum provided by stars via trapping of IR photons~\citep{Thompson2005} or by SN acquired during the Sedov-Taylor phase~\citep[e.g.,][]{Cioffi1988, Martizzi2015, Gatto2015, Kim2015}. 
The total momentum injection per unit stellar mass ($P_*/m_*$, used in the previous subsection) is related to $a_*(t)$ via
\begin{equation}
\label{eqn:new_pstar}
P_*/m_* = \int_0 ^\infty a_*(t) \cdot b \; dt
\end{equation}
where the boost factor $b$ is folded into the total momentum injection.
Setting the turbulent pressure from young stars equal to the mid-plane pressure given by equation (\ref{eqn:mid_plane_pressure}), we find the critical surface density of stars
\begin{equation}
\Sigma_{*,y}^{\rm crit} \approx \frac{2 \Sigma_g c_t \Omega }{f_p a_*(t)\cdot b } = \frac{ \pi G Q \Sigma_g ^2 }{ f_p a_*(t)\cdot b } .
\label{eqn:NewFeedbackEquilibrium}
\end{equation}
This result is similar to the argument in~\citet{Fall2010} and~\citet{Murray2010} for when individual star clusters disperse giant molecular clouds, 
but now we apply it to a galaxy nucleus which is allowed to have many {\it independent} self-gravitating clouds.

We briefly note that the same critical condition can be derived by considering the energy -- rather than pressure -- balance.
Specifically, expressing the turbulent energy injection rate as
\begin{equation}
\dot u \sim \frac{p_{*,y} c_t}{h} = \frac{ f_p a_*(t) \cdot b c_t \Sigma_{*,y} }{4   h }.
\label{eqn:NewFeedbackRate}
\end{equation}
and equating this with the turbulent energy dissipation rate given in eqn.~\ref{eqn:udot_diss} yields the same critical surface density of young stars.
This equivalence is built into the model with the assumption that the disk mid-plane pressure is related directly to the turbulent energy injection and dissipation rates.

Defining the young stellar fraction as
\begin{equation}
f_{*,y} \equiv \frac{ \Sigma_{*,y}  }{ \Sigma_{ {\rm tot} } }  
\end{equation}
we can write down the critical young stellar fraction or critical young star surface density in terms of the total surface density
\begin{equation}
f_{*,y}^{\rm crit} \equiv \frac{ \Sigma_{*,y}^{{\rm crit}}  }{ \Sigma_{ {\rm tot} } }   \sim  \frac{2 f_{\rm g} c_t \Omega   }{ f_p a_*(t)\cdot b }  \sim  \frac{ \pi G  Q f_g \Sigma_g  }{ f_p a_*(t)\cdot b}.
\end{equation}
If we assume a boost factor of $b=20$ and $a_*(t) = 1000 L_\odot / c M_\odot$ 
this reduces to
\begin{equation}
f_{*,y}^{\rm crit}  \approx 1.5 \times 10^{-2} f_g Q \left( \frac{\Sigma _g}{10^3 M_\odot \mathrm{pc}^{-2}} \right) .
\end{equation}
If this critical surface density of young stars is exactly maintained, feedback driven turbulence and disk self-gravity will be in equilibrium.
However, it follows that overshooting this critical young stellar surface density will lead to outflows, which will lower the system's gas fraction, and lead to a runaway outflow condition {\it that will persist for several local dynamical times}.\footnote{We caution that outflows can occur throughout our simulated disks even when this condition is not met.  The condition here will result in a strong, nearly complete, depletion of the local central gas reservoir whereas a steady level of gas outflow can be present elsewhere in the disk even when this condition is not met.}
For this reason, we argue that the critical surface density of young stars is an unstable critical point which will likely lead to a full shut down of star formation with the system going into an outflow state for a stellar evolution feedback timescale.  

After the young stellar populations have aged and their feedback levels have been reduced, gas will return to the nuclear region.
Gravitational torquing of gas via asymmetric disk potential features can drive gas mass influx rates of order $\dot M \sim -\left|a\right|^\beta \Sigma_{{\rm g}} R ^2 \Omega$
where $\left|a\right|$ is the asymmetric disk mode strength and $\beta=1-2$ depending on whether we are in the linear or non-linear torquing regime~\citep[see][for more details]{Hopkins2011c}.
This can repopulate the central gas reservoir within a few dynamical times, which will allow for a repetition of the central gas blowout cycle.

This result stands in contrast to the ``self-regulating" feedback balance that applies to most of the disk.  
The key difference is whether or not feedback driven turbulent pressure adjusts rapidly or slowly compared to the local dynamical time.
Typical galactic dynamical times at $\sim$kpc distances are of order $\sim100$Myrs (i.e. longer than the timescale over which stellar feedback evolves).
Upward perturbations to the SFR do not lead to blow-out under normal `equilibrium' conditions because stars die on timescales much shorter than the local dynamical time.
Where this is the case, there is not enough time for minor SFR perturbations to affect the galaxy dynamics.

\subsubsection{Equilibrium Breakdown from Insufficient Feedback}
Expressing the critical young star surface density as a fraction of the gas mass yields
\begin{equation}
\frac{ \Sigma_{*,y}^{{\rm crit}} }{ \Sigma_{\rm g}  }  = \frac{ 2 c_t \Omega    }{  f_p a_*(t)\cdot b} =   \Omega \frac{ Q \sigma f_g   }{   \sqrt{2} f_p a_*(t)\cdot b  }.
\end{equation}
An interesting limit is found by considering when the fraction of gas that must be turned into stars in order to achieve feedback balance is of order unity
(i.e. $ \Sigma_{*,y}^{{\rm crit}} / \Sigma_{\rm g}  \sim 1 $).  
This limit occurs for
\begin{equation}
 \Sigma_{\rm g}  = \frac{  f_p a_*(t)\cdot b }{  \pi Q  G    } \approx 2.7 \times 10^{11} \frac{M_\odot}{{\rm kpc}^2}
\end{equation}
for $b=20$ and $a_*(t)= 1000 L_\odot / c M_\odot$.
At this gas surface density, feedback can efficiently oppose gas collapse only by converting an order unity fraction of the gas mass into stars.
This critical surface density value approximately corresponds to the maximum observed stellar surface density in a wide range of dense stellar systems~\citep{Hopkins2010b} and 
represents a feedback failure mode that will be explored in more detail in a forthcoming paper (Grudic et al., in prep).

\section{Numerical Methods}
\label{sec:Simulations}
We investigate the galactic nuclear burst/quench cycles described above using numerical simulations.

\subsection{Simulation Code}
The simulations presented in this paper were performed using the N-body 
hydrodynamics code {\small GIZMO}~\citep{GIZMO}.  {\small GIZMO} 
is originally derived from {\small GADGET}~\citep{GADGET} and contains 
modifications centered around the ability to 
employ several fundamentally different hydro solvers.
In this paper, we employ {\small GIZMO} using the meshless-finite-mass method to solve the 
hydrodynamic equations of motion.

In addition to gravity and hydrodynamics, all simulations include a list of physical processes important to galaxy formation.
The specific galaxy formation model used in this paper is the FIRE feedback model~\citep{Hopkins2014}.
This model has been explored extensively in previous papers simulating the evolution of galaxies within cosmological volumes.
The FIRE model is capable of reproducing many observed galaxy properties including the stellar mass-halo mass relation, Kennicutt-Schmidt law~\citep{Hopkins2014},
the covering fractions of neutral hydrogen in the halos of z=2-3 Lyman Break Galaxies~\citep[LBGs,][]{FaucherGiguere2015}, 
and can self-consistently generate galactic winds that regulated galaxy mass growth consistent with observational requirements~\citep{Muratov2015}.

Details of the FIRE feedback model are given in~\citet{Hopkins2014}, which we summarize briefly here.
All simulations include radiative gas cooling and star formation with associated feedback.
Supermassive black hole particles are included in the present simulations because their gravity is dynamically important.
However, no black hole feedback is included.
Gas cools radiatively under the assumption of local thermodynamic equilibrium down to 10 K.
Dense/cold gas clouds are allowed to form stars if they are locally self-gravitating.
The star formation rate is given by $\dot \rho_* = \rho_{ {\rm mol }} / t_{ {\rm ff}}$ where the molecular fraction, $f_{{\rm H_2}}$, is inferred as a function of local gas column density and metallicity following~\citet{Krumholz2011}.  
Newly formed stars provide feedback to the ISM through thermal heating via supernovae, photo-ionization, local and long range radiation pressure, and stellar winds.
Each of these feedback sources has a well defined fiducial level that is adopted directly from {\small STARBURST99}
given a stellar particle's age and metallicity~\citep{SB99_3}.

\subsection{Initial Conditions}
We employ initial conditions that are modeled after redshift $z=0$ isolated star forming late type galaxies.
The initial distribution of particles is sampled based on the scaling relations of~\citet{MMWdisk} via the procedure outlined in~\citet{SpringelWhiteDisk}.
We construct three isolated galaxies that are similar in their physical characteristics but varied in initial mass to use as initial conditions.

The three galaxies have total masses of $M_{{\rm tot}}=\{1.39\times10^{11},3.37\times10^{11},1.39\times10^{12}\} M_\odot$.
Each of these galaxies consists of a dark matter halo, stellar disk, stellar bulge, gaseous disk (with 20\% gas fraction), and central supermassive black hole.
The three systems are similar in that they all contain 4 percent of their mass in the disk, a constant (small) fraction in the central supermassive black hole, 1 percent of their mass in the bulge, and the rest of the mass in the halo (see Table~\ref{table:ICProperties}).
The stellar and gaseous disks are given exponential surface density profiles with the same disk scale length set by the halo spin parameter of $\lambda=0.04$ ($R_d=\{1.7, 2.3, 3.7\}$ kpc).
The relative masses in stars and gas is set by the initial disk gas fraction, for which we adopt 20 percent.
The disks are initialized to be nearly in equilibrium (e.g., with $Q\sim1$).
The dark matter halo and stellar bulge are setup using a~\citet{H90} profile initial density distribution.
An NFW~\citep{NFW} equivalent concentration parameter of $C=10$ is used for all systems.
The bulge scale length is set to be 0.2 times the stellar and gaseous disk scale lengths.
The single supermassive black hole is included in each galaxy as a collisionless particle that is initially placed at the center of each galaxy.
Since the mass of the supermassive black hole is much larger than the typical gas or stellar particle mass, it remains close to the potential minimum of the host galaxy via self-consistently captured dynamical friction without any additional special treatment.
Additional values such as the force softening, mass resolution and naming convention that we used for this initial condition set are summarized in Table~\ref{table:ICProperties}.

Figure~\ref{fig:ICprofiles} shows the mass profiles, rotational velocity curves, and radially dependent dynamical time for our fiducial low mass galaxy (D1).  
The dynamical times are defined using the enclosed mass profiles to calculate the circular velocity.
Although we neglect black hole accretion and feedback in these sets of simulations, the presence of this massive particle is important owing to its impact on the dynamics of the nuclear region.
The presence of the central supermassive black hole dominates the mass enclosed and rotation curve for the inner 10-30 parsecs.  
The rotational velocity curves have a non-monotonic profile, but the dynamical time continually drops toward the galaxy center.
The characteristic stellar evolution timescales of $\sim 3$ Myr and $\sim 40$ Myr are identified with horizontal dashed lines in the right panel of Figure~\ref{fig:ICprofiles}.
We identify $3$ and $40$ Myr lines because this roughly corresponds to the time when stars begin moving off the main sequence and when the most of the energy from the newly formed stellar population has been deposited, respectively.
For this setup the dynamical time drops below $\sim 3$ Myr at a radius of $\sim 100$ pc, which changes only slightly for the other initial disks.
Within this region, we expect star formation and gas supply to be unable to achieve a steady state solution owing to the short dynamical times.
The presence of a $10^7 M_\odot$ black hole alone keeps the dynamical time below $3$ Myr for the central $\sim 100$ pc, but a massive/compact stellar bulge could play a similar role in shaping the inner galactic potential.

\begin{figure}
\centerline{\vbox{\hbox{
\includegraphics[width=0.475\textwidth]{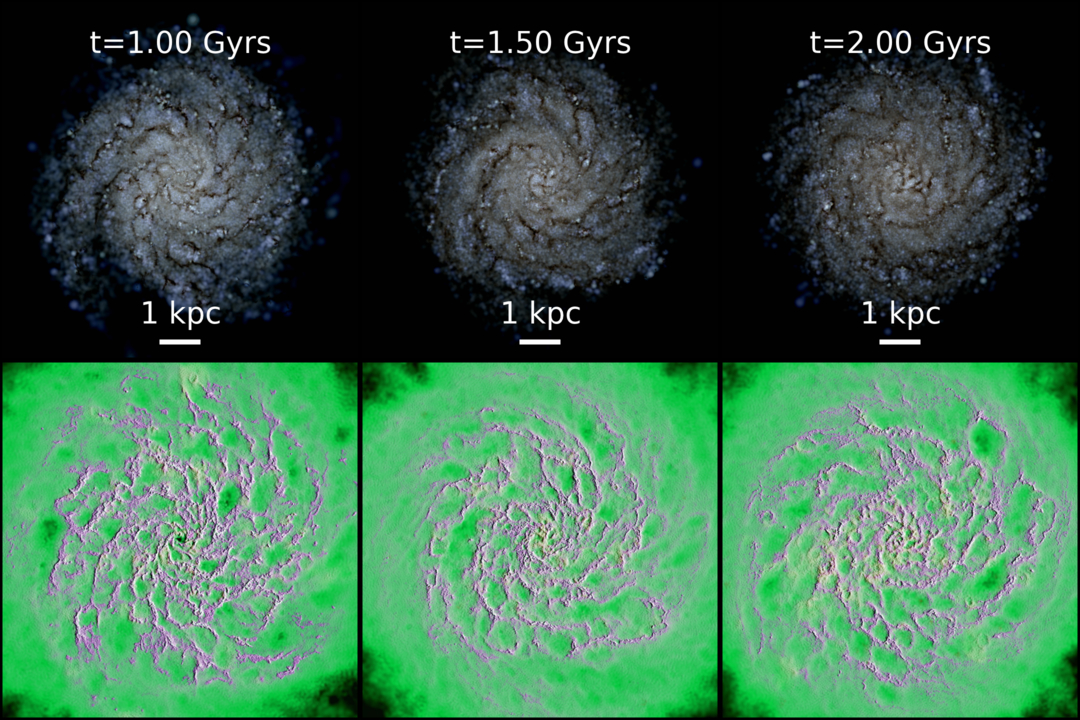}}}}
\caption{The projected gas surface density is shown at three snapshots during the evolution of the D1 system.  
The top row shows synthetic SDSS-g, -r, -i band images made using a simple line of sight attenuation model.
The bottom row shows the gas surface density, with magenta indicating atomic/molecular gas ($10$ K$<T<8\times10^3$ K), green indicating warm ionized gas ($8 \times 10^3$ K$< T<8 \times10^5$ K), and yellow (though little is present) hot gas ($T> 10^5$ K).
The time of each snapshot is marked in the top row and the scale bar of 1 kpc is also indicated.
From the initially smooth gas distribution a turbulent ISM develops, with dense star forming regions being continually formed by self-gravity and destroyed through feedback from young stars.}
\label{fig:gas_surface_density_1}
\end{figure}

\begin{table}
\begin{center}
\caption{The properties of the initial conditions used for the three isolated disks presented in this paper.  
Below, $M_{tot}$ is the total mass of each system and $m_p$ is the baryon particle mass. 
Adaptive gravitational softening lengths are employed with $\epsilon$ being the minimum (Plummer equivalent) force softening length used for each particle type.
}
\label{table:ICProperties}
\begin{tabular}{ l l c c c c c }
ID 	&							&	Gas 	Disk			& Halo   				& Stellar  	Disk			\\	
	&							&					&					& 					\\
\hline	
\hline 
D1  	&	M${}_{{\rm tot}} \;[M_\odot ]$ 	& $1.11 \times 10^9 $	&   $1.32 \times 10^{11}$ 	&  $4.45 \times 10^9$  	\\
	&	m${}_{{\rm p}} \;[M_\odot ]$& $10^3  $				&   $8.0 \times 10^4$  	&  $3.0 \times 10^3$  	\\
  	&	$\epsilon \;[pc]$				& 1   					&   10  				&  2   				\\
\hline
D2  	&	M${}_{{\rm tot}} \;[M_\odot ]$ 	& $2.72 \times 10^9 $	&   $3.23 \times 10^{11}$  	&  $1.09 \times 10^{10}$  	\\
	&	m${}_{{\rm p}} \;[M_\odot ]$ 	& $10^3 $				&   $8.0 \times 10^4$  	&  $3.0 \times 10^3$  	\\
  	&	$\epsilon \; [pc]$			& 1   					&   10  				&  2   				\\
\hline
D3  	&	M${}_{{\rm tot}} \;[M_\odot ]$	& $1.11 \times 10^{10} $	&   $1.32 \times 10^{12}$  &  $4.45 \times 10^{10}$  	\\
	&	m${}_{{\rm p}} \;[M_\odot ]$	& $10^3 $				&   $8.0 \times 10^4$  	&  $3.0 \times 10^3$  	\\
  	&	$\epsilon \;[pc]$				& 1   					&   10  				&  2   				\\
\hline
\hline
\end{tabular}
\end{center}
\end{table}

\section{Results}
\label{sec:Results}
\subsection{Gas Distribution and Galaxy Structure}
Figure~\ref{fig:gas_surface_density_1} shows a time sequence of the D1 galaxy evolved from 1 to 2 Gyr. 
The top panel shows a synthetic optical image of the galaxy while the bottom panel shows the gas content. 
As indicated in both panels, throughout the entire simulation the disk remains globally stable without any strong bars developing (although there is a prominent spiral pattern).
This global stability is by construction.
As the simulation proceeds gas cools and collects into dense star forming clumps which are embedded in a volume filling warm phase of the ISM. 
The color of the gas distribution on the bottom panel highlights this multi-phase structure: magenta gas indicates atomic/molecular material while green indicates warm ionized gas. 
These star forming clumps are eventually disrupted through feedback from young stars.
The self-regulation of a multi-phase ISM through feedback from stars using the FIRE feedback model has been discussed extensively in~\citet{Hopkins2014}.
Although we only show results here for the D1 disk, we note that D2, and D3 show qualitatively indistinguishable behavior.

\begin{figure*}
\centerline{\vbox{\hbox{
\includegraphics[width=1.0\textwidth]{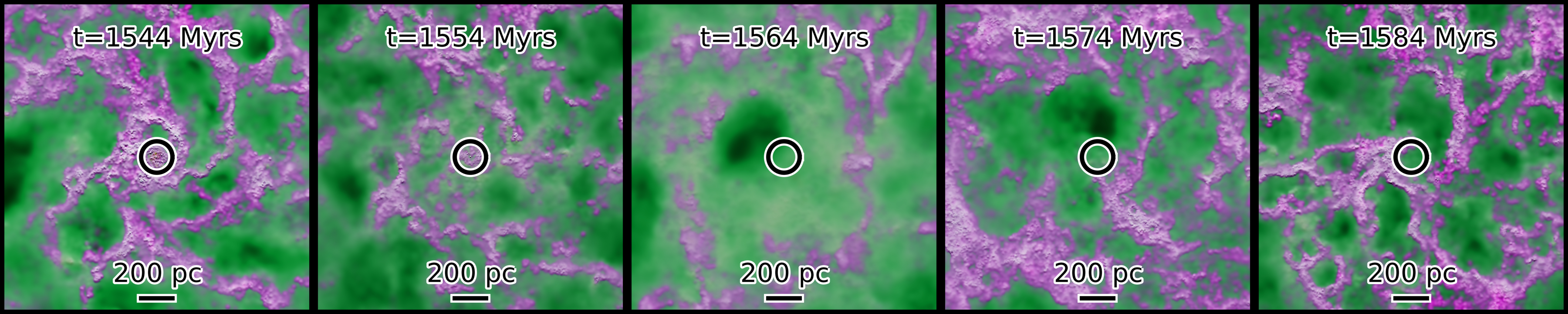}}}}
\caption{The projected gas surface density is shown at five snapshots during the evolution of the D1 system, focused around the galactic nucleus during a period of nuclear gas blowout.  
The white circle indicates a cylindrical aperture of 100 pc centered on the central supermassive black hole, which stays closely pinned to the potential minimum.
Significant molecular gas is initially present in the central 100 pc aperture.  The concentration of cold/molecular gas (indicated with purple) in the nuclear region in the first (leftmost) snapshot  has an associated elevated level of nuclear star formation.
Soon after the feedback from young stars hits the critical level, leading to a near complete blowout of gas within the central 100 pc.
The gas density remains suppressed for the next $\sim$30 Myr, while the feedback from young stars continues to be efficient in spite of the absence of much ongoing star formation.}
\label{fig:gas_surface_density_2}
\end{figure*}

The multi-phase ISM stretches across the full radial range of each galaxy, including the central region.  
To highlight the evolution of the ISM phase structure in the nuclear region, Figure~\ref{fig:gas_surface_density_2} shows the gas surface density within a narrow field of view focused on the galactic nucleus at several times.
We can identify here the ISM structure extending down to $\ltsim$100 pc scales.
The gas density in the central region can be seen to fluctuate through the images.
The most severe gas fluctuations are short lived and so we have selected times that highlight a period of central gas blowout. 
The typical blowout cycle begins with a concentration of molecular (purple) star forming gas in the central $\sim 100$ pc, as is found in the left most panel.
This dense gas concentration forms stars rapidly until the feedback from these young stars reaches the critical $f_{*,y}^{\mathrm{crit}}$ value.
This transition occurs very rapidly as the dynamical times on these scales is only $\sim$1 Myr (see Figure~\ref{fig:ICprofiles}).  
After gas is removed from the central region, the gas remains at very low densities for $\sim$20-30 Myr (i.e. the third and fourth panels).
By $\sim$50 Myr (the rightmost panel) molecular gas begins to build up again just outside of the central region.
Asymmetric disk features are clearly present which foster the torquing of gas to the central 100 pc.

\begin{figure*}
\centerline{\vbox{\hbox{
\includegraphics[width=1.0\textwidth]{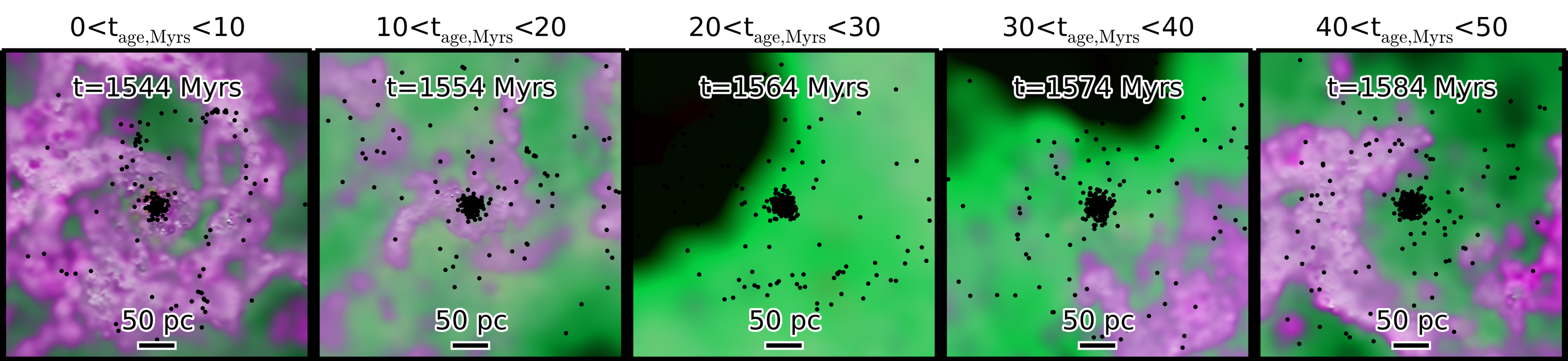}}}}
\caption{Same as Figure~\ref{fig:gas_surface_density_2}, but zoomed further on the galaxy nucleus and with the locations of young star particles identified with black circles.
The left panel indicates the locations of the youngest stars which are highly clustered and generally coincide with the dense gas in which they were formed.
Subsequent panels show the same star particles as they age.
By a few tens of millions of years the distribution of young stars becomes more randomized in this region with the stellar populations being well separated from their birth clouds.
This is partially driven by feedback, but aided significantly by the short dynamical times on these scales.  }
\label{fig:young_star_dist}
\end{figure*}

In addition to considering the gas distribution on these scales, we consider the distribution of young star clusters and their evolution with time.
Figure~\ref{fig:young_star_dist} shows the evolution of the stellar populations associated with this same blowout event.
Specifically, the stellar population that was less than $10$ Myr old is identified in the leftmost panel, and those same particles are tracked forward in time. 
In the leftmost panel, the young stars (here $t<10$ Myr) take on a clustered distribution that is (mostly) coincident with dense star forming gas.
In some cases, young stars are found in low density regions when they have already been able to blow apart their birth cloud.
Within $10$ Myr (second panel from left), the stars have already randomized their locations significantly.  
There is little correlation between these young stars and the dense gas distribution.
By $20$ Myr (central panel), the stars continue to deposit significant momentum and energy via supernovae but the gas is almost completely evacuated.
However, once the feedback from this young stellar population recedes (two rightmost panels) dense/cold molecular gas quickly returns to the central region.

From Figure~\ref{fig:young_star_dist} we identify that star formation occurs in several distinct self-bound dense gas clouds within the central $\sim$100 pc.  
However, the young stars rapidly decouple from their birth clouds owing both to feedback and the short dynamical time.
The rapid mixing of the young stellar populations drives the whole central region to react homogeneously to the presence of strong feedback.
Despite the presence of multiple self-bound star forming clumps, the feedback response of the gas in the whole nuclear region acts like that of a single molecular cloud:
stars form until a critical surface density of young stars is present (such that feedback overwhelms gravity) at which point they begin to expel gas from the nuclei.

\subsection{Star Formation Rates and Gas Masses Over Time}
\begin{figure*}
\centerline{\vbox{\hbox{
\includegraphics[width=7.0in]{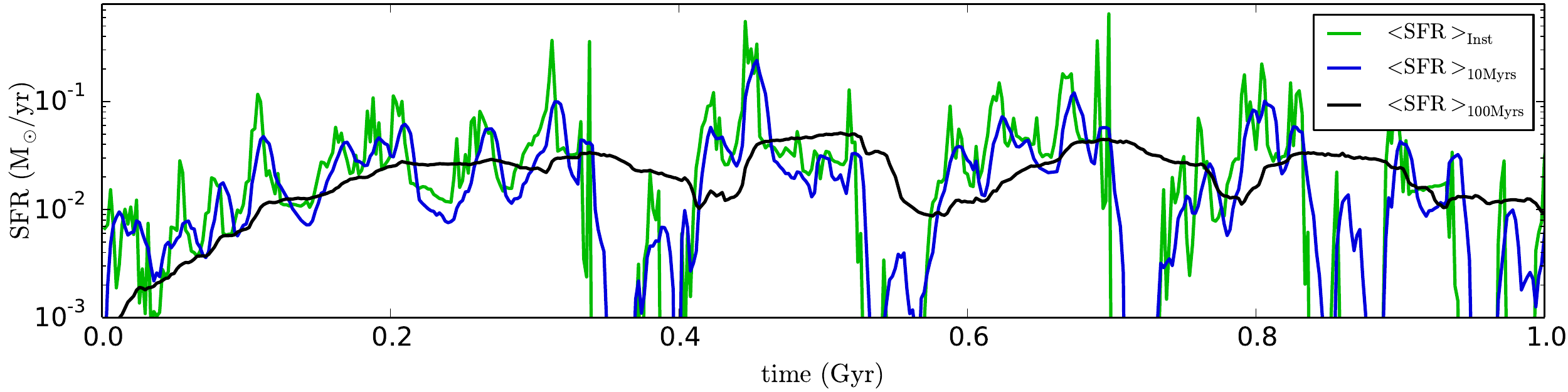}}}}
\caption{The SFR within the central $100$ pc as a function of time for the D1 disk using the instantaneous SFR (green line), the mass of stars formed over the past 10 Myr (blue line), and the mass of stars formed over the past 100 Myr (black line).  
The mass of young stars gives a smoothed estimate of the SFR and provides a better match to observational SFR probes.  
We exclusively use the mass of young stars formed in the past 10 Myr (blue line) to determine the SFR throughout the rest of this paper.  }
\label{fig:smoothed_sfr_rate}
\end{figure*}

Figure~\ref{fig:smoothed_sfr_rate} shows the SFR within the central 100 pc for the D1 galaxy.
We show the SFR when taken as an instantaneous value from the simulation, as well as averaged over the past 10 and 100 Myr by calculating $\mathrm{ SFR} = M_*(t<t_0) / t_0$ where $t$ is the age of the star and $t_0$ is the averaging timescale.  
We present smoothed definitions of the SFR for two reasons.  
First, the instantaneous SFR values obtained from the simulations rapidly vary and with significant magnitude.
Smoothing the SFRs reduces the very high frequency noise allowing for more clear identification of the mean evolutionary trends.
Second, observational SFR indicators are not sensitive to the instantaneous SFR, but rather the total mass of young stars formed over some recent time window.
The smoothed definition of the SFR is therefore more in line with what is measured observationally through UV or nebular emission line SFR tracers~\citep[e.g., see Table 1 of][]{Kennicutt2012}.
In the rest of this paper, we adopt a definition for the star formation rate as the average star formation rate over the past 10 Myr -- which would be most similar to what would be measured though H-alpha SFR measurements.  

\begin{figure*}
\centerline{\vbox{\hbox{
\includegraphics[width=7.0in]{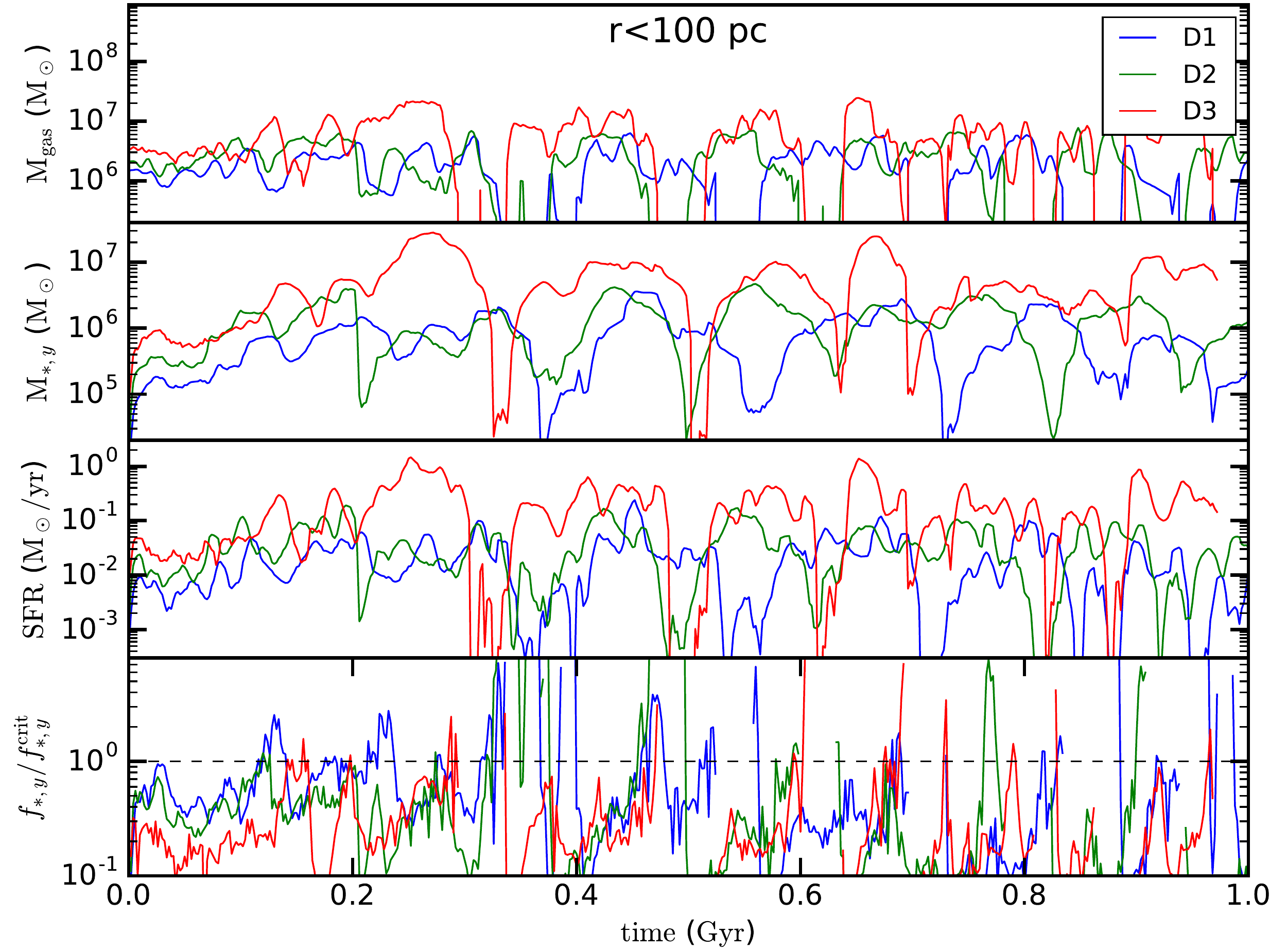}}}}
\caption{Top to bottom: The gas mass, young stellar mass, star formation rates, and critical young stellar fraction as a function of time for the central 100 pc.
The episodic star formation in the central 100 pc is clear.  
Gas content first builds up in the central region, then star formation rapidly builds a population of young stars, which drive sufficiently strong feedback to drive down the central gas density. }
\label{fig:gas_nuclear_evolution}
\end{figure*}

\begin{figure*}
\centerline{\vbox{\hbox{
\includegraphics[width=7.0in]{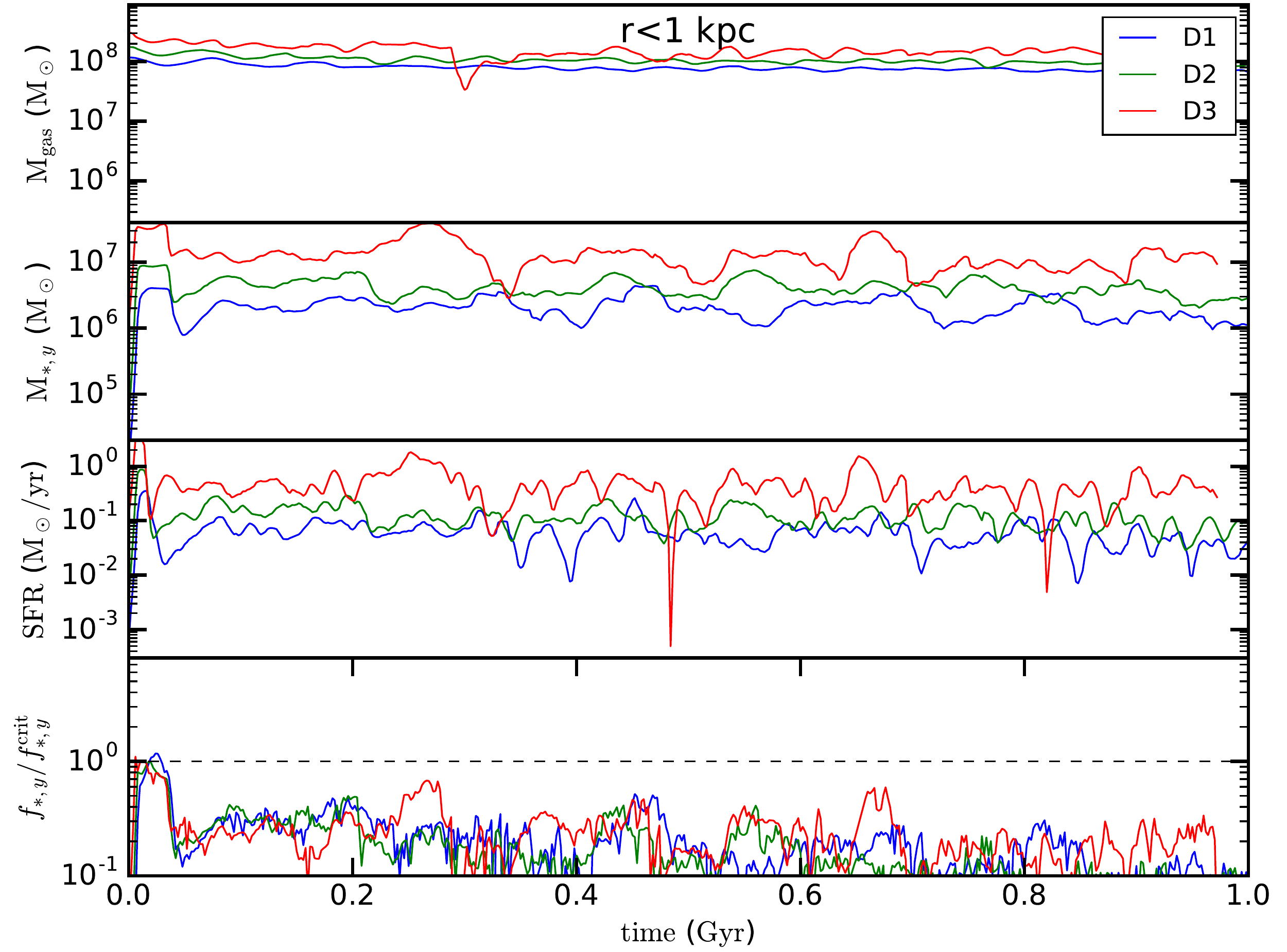}}}}
\caption{Same as Figure~\ref{fig:gas_nuclear_evolution} but for the central 1 kpc.  
The oscillatory properties that were present for the central 100 pc are not evident for the central 1 kpc.
All measured quantities are stable over this larger aperture, in contrast to the 100 pc central region.}
\label{fig:gas_nuclear_evolution2}
\end{figure*}

The episodic central star formation activity is directly examined in Figure~\ref{fig:gas_nuclear_evolution} for the simulated systems. 
The panels of Figure~\ref{fig:gas_nuclear_evolution} show the time evolution of the gas mass, young stellar mass (i.e. $M_*(t<40$Myr$)$), star formation rate, and critical feedback value from top to bottom.
Figure~\ref{fig:gas_nuclear_evolution} presents these quantities within the central 100 pc, while Figure~\ref{fig:gas_nuclear_evolution2} shows the same quantities within the central 1 kpc.

Over very long periods of time (i.e. $\sim$ 1 Gyr timescales) we find the central region's gas mass and star formation rate is characterized by a relatively steady state.
The time averaged star formation rate or central gas mass evolves only marginally over the full time period shown in Figure~\ref{fig:gas_nuclear_evolution}.
This is impressive, given that in the nuclear regions of the galaxies explored here are being evolved for hundreds of dynamical times.
The steady nature of the SFR over $\sim$1 Gyr time scales is consistent with star formation being effectively regulated by stellar feedback. 
Presumably, however, this would break down on time scales much greater than $\sim$1 Gyr, where the absence of IGM accretion would lead to disk gas depletion.  

On shorter timescales the central 100 pc region is characterized by an oscillatory gas mass, star formation rate, and young stellar mass within the nuclear region.
The behavior of each of these components is offset in time but closely related.
The central gas mass increases in the absence of current strong stellar feedback until star formation becomes efficient in the central region owing to the presence of sufficiently dense, self-bound gas clumps.
Star formation then proceeds efficiently until stellar feedback is sufficiently strong as defined by equation~\ref{eqn:NewFeedbackEquilibrium} to balance gas collapse.
Once feedback is sufficiently strong, the gas supply is rapidly expelled (characterized by very sharp drops in the central gas content for any of the disks).
The young stars that are responsible for the strong feedback continue to inject strong feedback into the nuclear region for $\sim$30-40 Myr causing the gas recovery in this region to be relatively slow.
The episodic sharp drops in the central gas content are correlated with the periods of peak young stellar mass.
The timescale for this behavior is of order $\sim$100 Myr -- though the detailed duty cycle for this process is not very regular, and is different for our three disk models.
The time evolution shown in Figure~\ref{fig:gas_nuclear_evolution} can be compared back against the central gas surface density plots shown in Figure~\ref{fig:gas_surface_density_2}.  
Both figures present a complementary picture of the cyclical gas blowout from the central region.

The behavior of the same quantities averaged on larger spatial scales becomes significantly more stable.
Examining Figure~\ref{fig:gas_nuclear_evolution2} we find that averaged over the central 1 kpc, the mass of gas and young stars in the central regions shows no clear oscillatory behavior. 
The instantaneous star formation rate (not shown) shows significant variability, which becomes nearly smooth when averaged over 10 Myr periods (the plotted curves).
The increased stability on 1 kpc scales is consistent with the analytic arguments outlined in Section~\ref{sec:AA}.
Although feedback operates to disrupt gas clouds locally feedback failed to collectively clear 
material from the region.
The rate of gas consumption via star formation and gas expulsion via feedback are well balanced with the rate of gas return from aging stellar populations and influx from larger radii producing
stable gas mass evolution.

\subsection{The Threshold for Burst/Quench Cycles} 
In Section~\ref{sec:AA} we derived the surface density of young stars that were required to drive the outflows.
The surface density of young stars normalized by the critical surface density that is expressed in Equation~\ref{eqn:NewFeedbackEquilibrium} is plotted as a function of time in the bottom panels of Figures~\ref{fig:gas_nuclear_evolution} and~\ref{fig:gas_nuclear_evolution2} for our three disks as evaluated within the central 100 pc and 1 kpc, respectively.
Periods of time with $f_{*,y}^{\rm crit}<1$ indicate that the system has low levels of feedback compared to what is required for gas blowout and is therefore stable/collapsing while 
periods of time with $f_{*,y}^{\rm crit}>1$ indicate that the system is feedback dominated and is likely driving outflows.  
The central portion of each galaxy spends the majority of its time in a stable/collapsing state, bookended with brief periods of strong feedback. 
We can compare the periods of time with high $f_{*,y}^{\rm crit}$ values with the time evolution of the quantities presented in the bottom panel of Figure~\ref{fig:gas_nuclear_evolution}.
When $f_{*,y}^{\rm crit}$ exceeds unity, there is a strong correspondence to a rapid drop in the central gas content of the system.
Contrasting the panels within Figure~\ref{fig:gas_nuclear_evolution} reveals that the $f_{*,y}^{\rm crit}$ values rise much more sharply than the young stellar mass.
The driving force behind this feature is that while the mass of young stars is initially responsible for pushing $f_{*,y}^{\rm crit}$ above unity, 
once this is achieved the rapid outflow of gas mass from the central region drives up the derived turbulent velocity and drives down the orbital frequency and gas mass.
The combination of these three effects results in the very sharp rise of the $f_{*,y}^{\rm crit}$ values.  
Very large $f_{*,y}^{\rm crit}$ values are not long lived because (i) the gas fraction drops, (ii) new star formation ceases, and soon after (iii) the mass of young stellar mass decreases (owing to aging).

\begin{figure*}
\centerline{\vbox{\hbox{\includegraphics[width=7.0in]{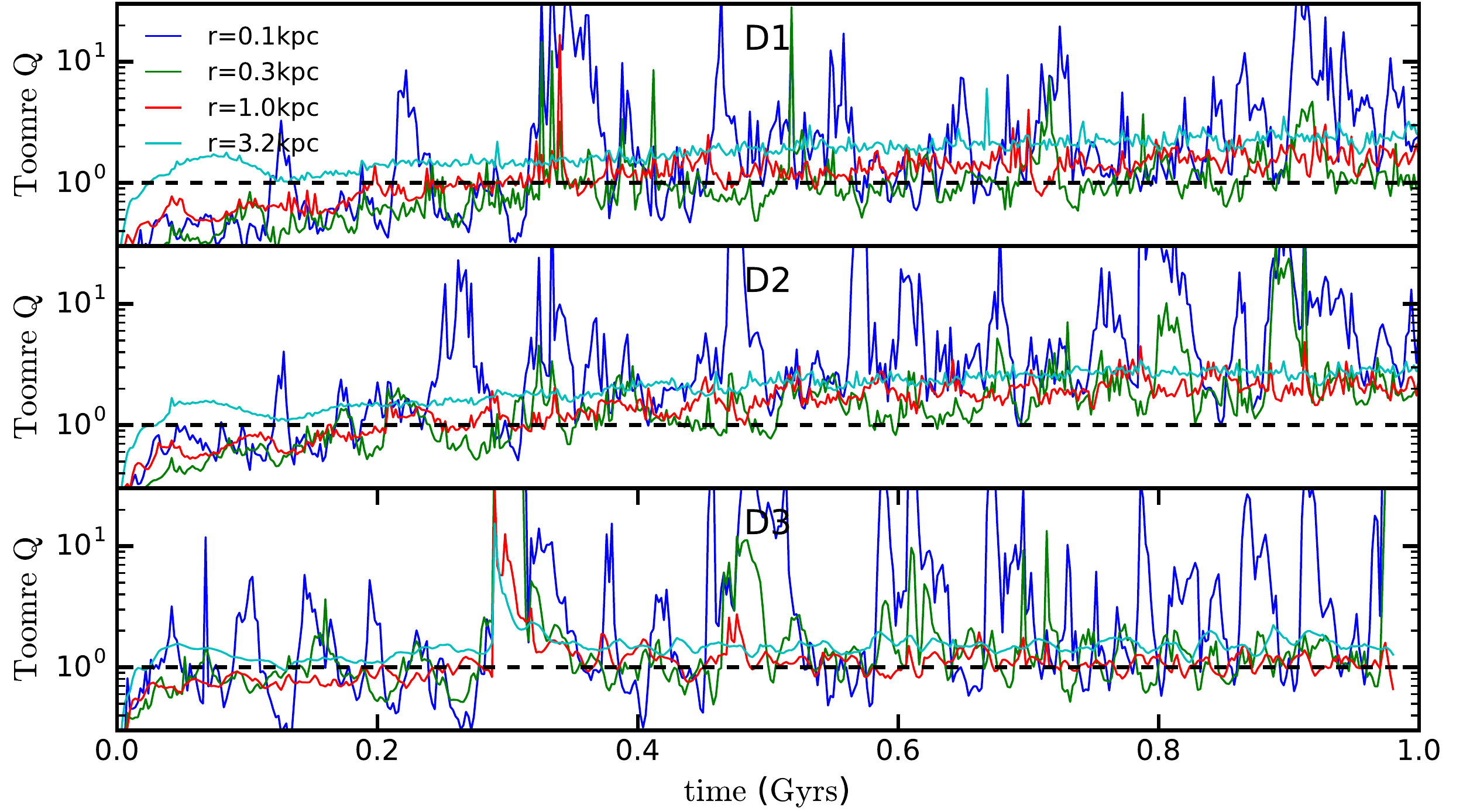}}}}
\caption{(Left) The Toomre Q parameter is shown as a function of time for several radii within each disk.  
Q approaches unity for all radii.  
However, while Q  hovers around unity consistently with time for the large radii bins, the smaller radii bins have episodic and strong increases in the Toomre Q value.  
These increases in Q correspond to periods where stellar feedback temporarily dominates the central region's dynamics, and are indicative of local outflows.
(Right) The $r, \phi, z$ velocity dispersion is shown as a function of time for each disk for the central r=100 pc.
Steady state velocity dispersion of $c_t\sim$10-50 km/sec are achieved, with brief excursions to higher velocity dispersion when outflows are being launched.
}
\label{fig:toomre_q}
\end{figure*}

\subsection{Toomre Q and Gas Velocity Dispersions}
Further insight into the dynamic behavior of the central gas reservoir can be obtained from Figure~\ref{fig:toomre_q} which shows the Q parameter as a function of time for several disk radii.  
The Q value is evaluated via $Q = 2 c_t \Omega / \pi G \Sigma_{\rm g}$ using the $\hat z$ component of the mass weighted gas velocity dispersion as a proxy for the turbulent velocity dispersion.
We adopt the vertical, mass weighted, velocity dispersion because the radial and azimuthal velocity dispersions are more subject to influence by the coherent, non-circular motion of dense clouds in the disk plane.
Moreover, the mass (rather than volume) weighted velocity dispersion is both more important for the disk dynamics, and less subject to influence from rapidly outflowing material.
The Q value hovers around unity consistently with time at large radii, indicating steady-state self-regulation.  
The smaller radii bins have episodic and strong increases in the Toomre Q value, indicative of local outflows.  

The inner regions of the simulated disks experience brief periods of large gas velocity dispersions coincident with the presence of locally-driven outflows.
Between these periods, the gas velocity dispersions are of order 30-50 km/sec.
More specifically, the D1, D2, and D3 disks have median vertical gas velocity dispersions of in the central 1 kpc of $35.1$, $46.3$, and $30.7$ km/s, respectively.
This is consistent with the prediction that $c_t=f_g \sigma / 2$ for $Q=1$~\citep[our $c_t$ is slightly higher owing to the multi-component disk and order unity corrections calculated in][]{FaucherGiguere2013}.
The gas velocity dispersion measured on smaller scales is similar in magnitude (i.e. 30-50 km/sec), but with significantly more variability -- particularly around burst episodes.

\subsection{Nuclear Kennicutt-Schmidt Relation}
\begin{figure*}
\centerline{\vbox{\hbox{
\includegraphics[width=3.3in]{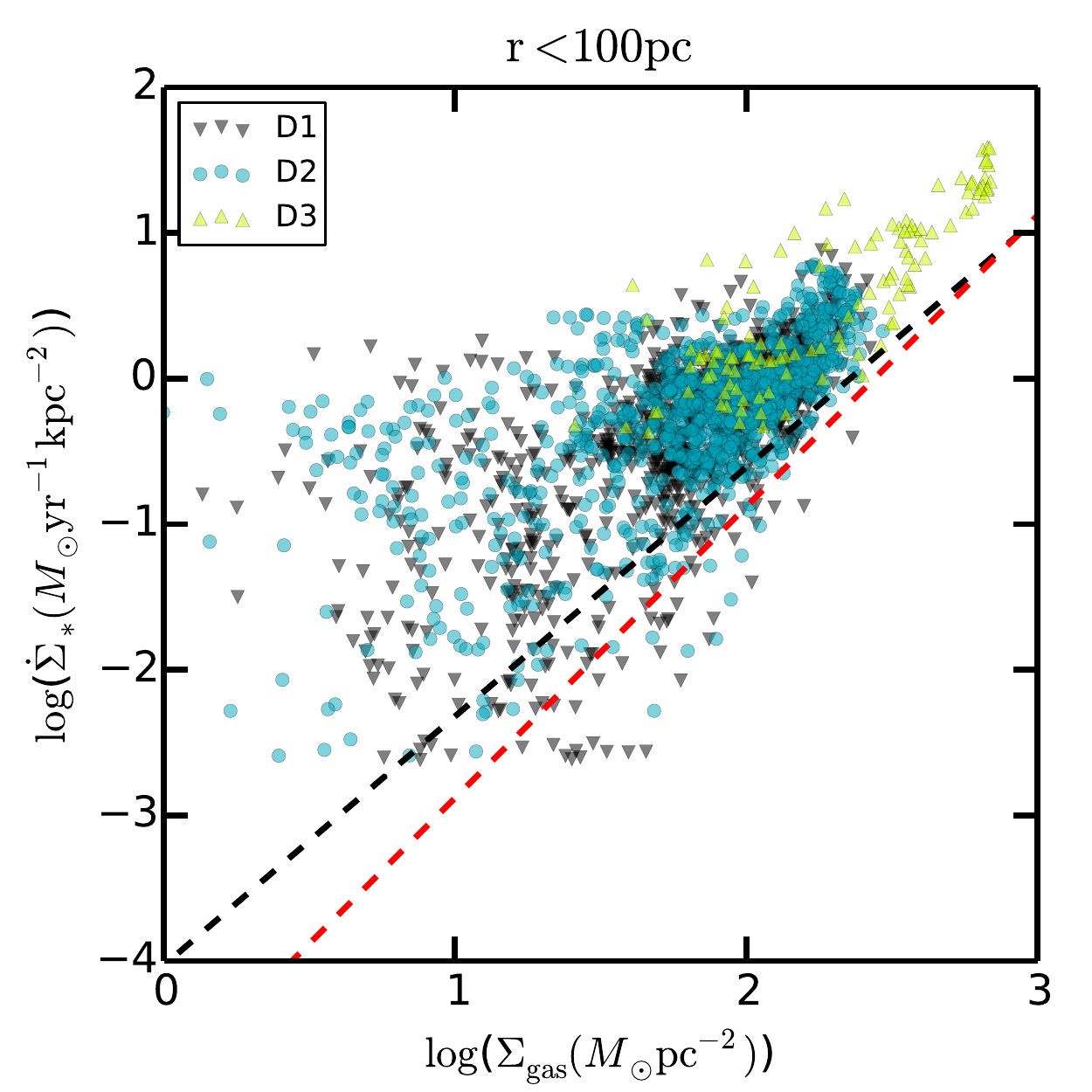}
\includegraphics[width=3.3in]{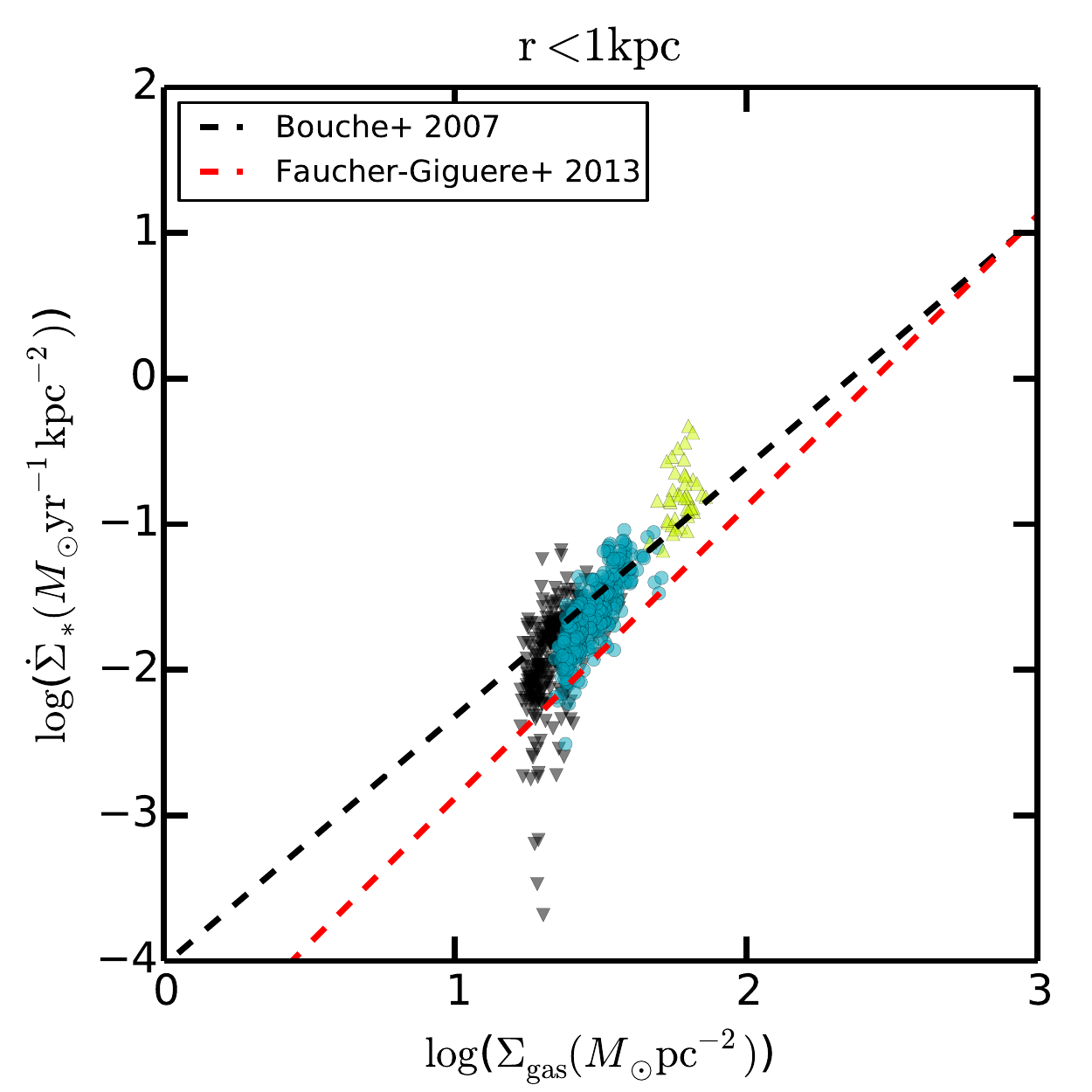}
}}}
\caption{The $\Sigma_{{\rm SFR}}$-$\Sigma_{{\rm gas}}$ KS relation is shown for material within the central r=$100$ pc (left) and r=$1$ kpc (right).  
We find significantly increased scatter for the r=$100$ pc  relation compared to larger aperture relation. 
For comparison, we present the empirically derived KS relations from~\citet{FaucherGiguere2013} and~\citet{Bouche2007}.   }
\label{fig:ks_relation1}

\centerline{\vbox{\hbox{
\includegraphics[width=3.3in]{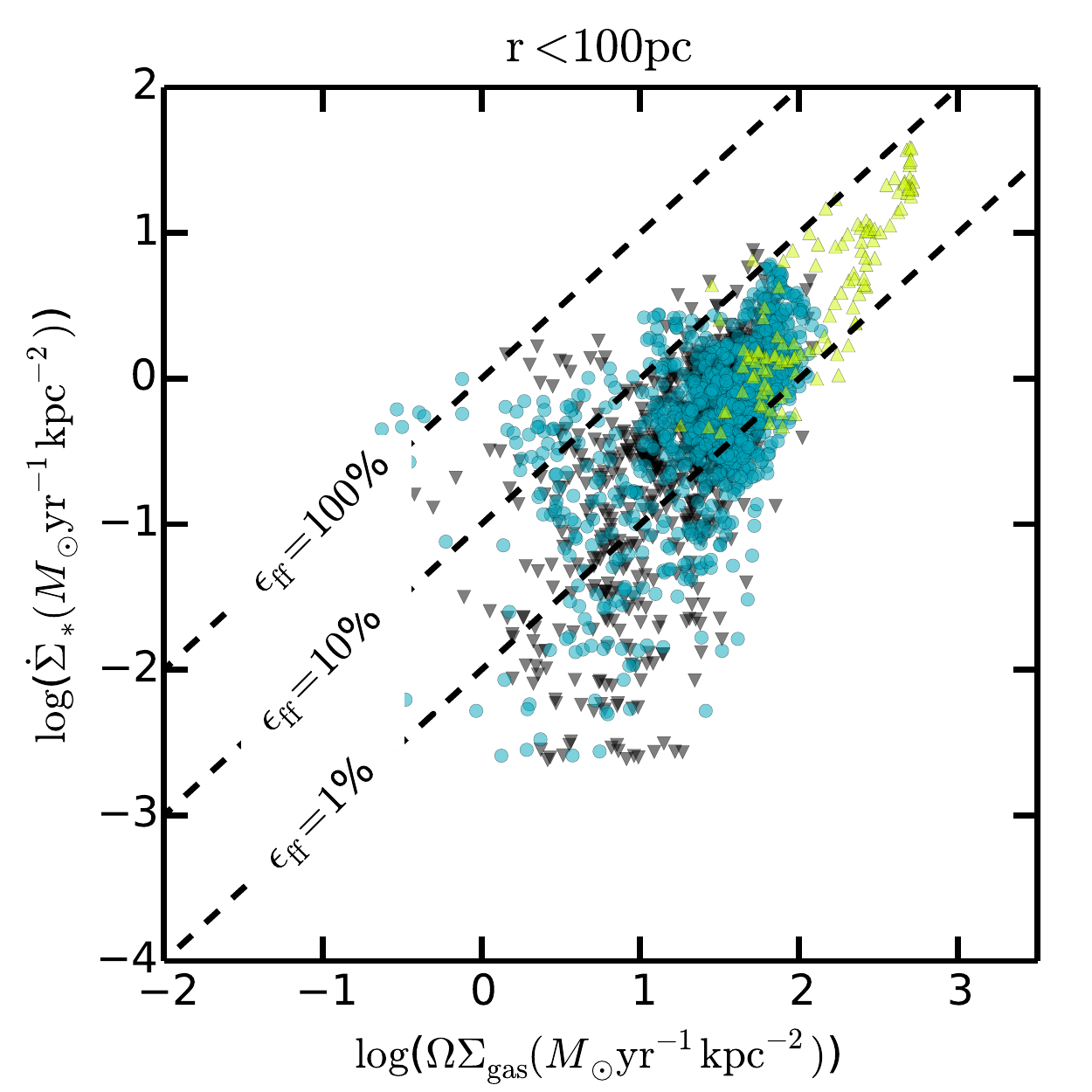}
\includegraphics[width=3.3in]{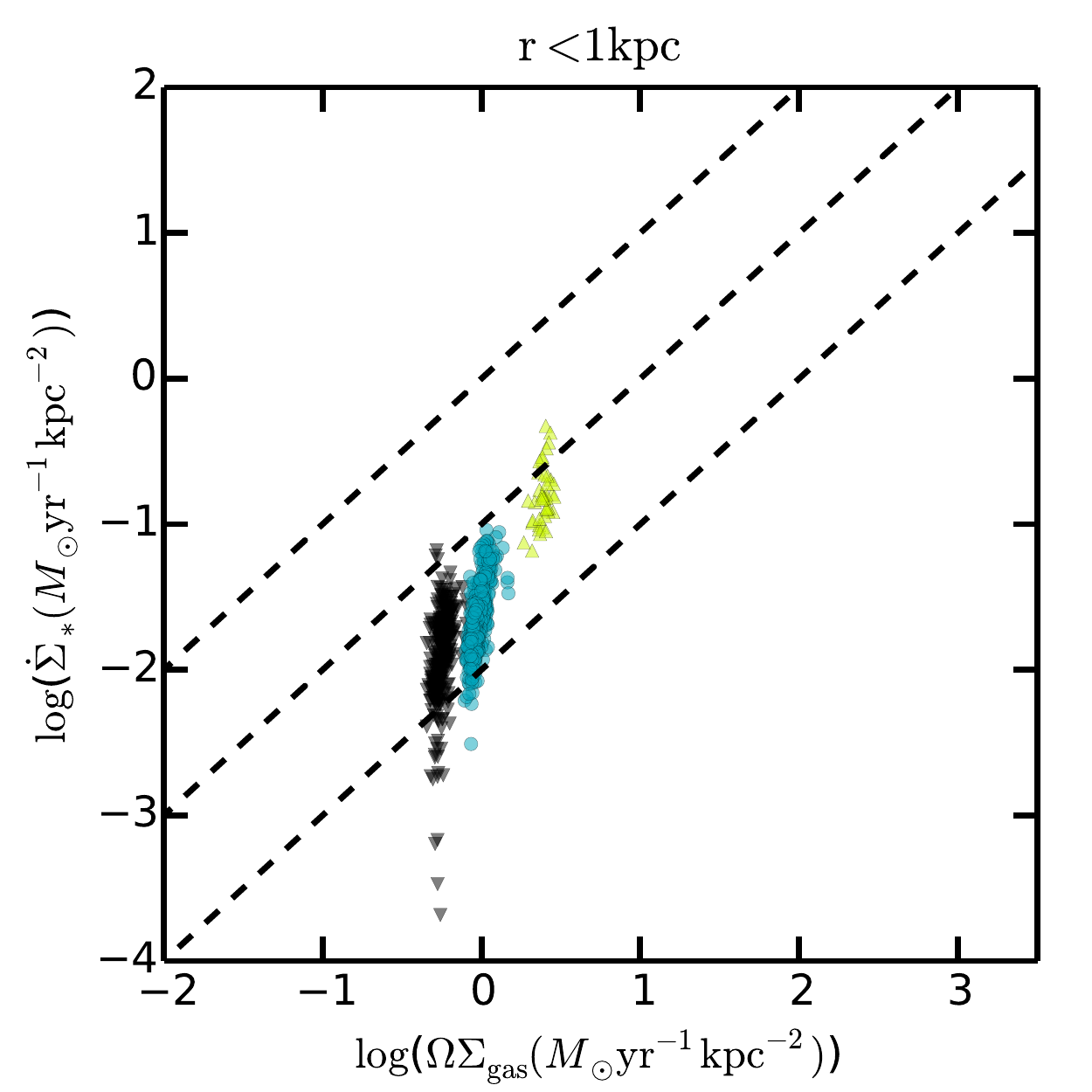}
}}}
\caption{The $\Sigma_{{\rm SFR}}$-$\Omega\Sigma_{{\rm gas}}$ relation for material within the central r=$100$ pc (left) and r=$1$ kpc (right) is shown. 
The same legend from Figure~\ref{fig:ks_relation1} applies.
The dashed lines in the plots indicate 1\%, 10\%, and 100\% of the gas mass being turned into stellar mass per dynamical time.
The star formation efficiency in the nucleus has more scatter than that at larger radii. 
}
\label{fig:ks_relation1.5}
\end{figure*}

Equilibrium analytic models have been used to explore the low efficiency of star formation that is observed in the KS relation~\citep{Thompson2005, Ostriker2011, FaucherGiguere2013}.
Their breakdown may therefore have implications for the KS relation in those regions.
In Figure~\ref{fig:ks_relation1} we consider the $\Sigma_{{\rm SFR}}$-$\Sigma_{{\rm gas}}$ KS relation for material within the central r=$100$ pc (left) and r=$1$ kpc (right).
Figure~\ref{fig:ks_relation1.5} shows the $\Sigma_{{\rm SFR}}$-$\Omega\Sigma_{{\rm gas}}$ relation -- again for material within the central r=$100$ pc (left) and r=$1$ kpc (right).\footnote{For clarity: we take the dynamical time to be $t_{{\rm dyn}}=1/\Omega = r/v$.  We calculate the dynamical time based on the local potential.}
Star formation rate surface densities are calculated based on the mass of stars less than 10 Myr old -- consistent with the definition of SFR used throughout the rest of this paper.

The KS relation averaged over the central $\sim$1 kpc is reasonably tight and consistent with the 
KS relations derived from observations by~\citet{Bouche2007} and ~\citet{FaucherGiguere2013}. 
The slope of the KS relation predicted by our simulations is roughly n$\sim$2 ($\Sigma_{\mathrm{SFR}} \propto \Sigma_g^n$). 
This slope is steeper than slopes n$\sim$1-1.5 often derived from observations assuming a constant $X_{\mathrm{CO}}$ conversion factor~\citep[e.g.,][]{Genzel2010} but is consistent with analytic equilibrium models~\citep{Ostriker2011, FaucherGiguere2013}, 
and with an interpretation of observations assuming an $X_{\mathrm{CO}}$ factor varying continuously with $\Sigma_g$ \citep{Ostriker2011, Narayanan2012}.
This has been shown already for our adopted ISM/star-formation/feedback model~\citep{Hopkins2011a, FaucherGiguere2012, Hopkins2015}.
The SF efficiency for all three disks hovers around 5\% star formation efficiency per dynamical time.

The smaller aperture KS relations have significantly more scatter than their larger aperture counterparts.
The dense cloud of points sitting around $\Sigma_{\mathrm{gas}} \sim 10^2 M_\odot \mathrm{pc}^{-2}$ falls somewhat above the observed KS relation.
This is owing to the $100\%$ local SF efficiency per free fall time (at the star formation density threshold) that is used in our simulations.
Once stars are formed in this cloud, they immediately start acting to clear the nuclear region of gas.
Since the SFRs plotted in Figures~\ref{fig:ks_relation1} and~\ref{fig:ks_relation1.5} are derived from the mass of young stars we find a cloud of points that extends toward significantly lower gas surface densities.
This is a result of measuring the SFR from the mass of young stars that have both formed recently and begun to clear their birth cloud gas, and comparing it against the current gas mass.
Reexamination of Figure~\ref{fig:young_star_dist} confirms how this process takes place in our simulations.

\section{Discussion}
\label{sec:Discussion}

We have examined the stability of self-regulated star formation in galactic nuclei.  
Star formation in regions of galaxies with short dynamical times (shorter than the stellar evolution timescale) cannot reach a truly equilibrium state, but instead undergo ``burst quench" cycles.
We presented an analytically motivated model where gas flows into the nuclear region and begins forming stars 
until enough young stellar mass is present to overwhelm the self gravity of the gas leading to a localized blowout of the gas.

This is similar to the behavior of a single GMC~\citep[see, e.g.,][]{Fall2010, Murray2010} within which the local dynamical time can also become short.
The key difference between the nuclear region and a typical GMC is that this behavior is driven by the collectively short dynamical time of the disk, rather than the local potential of the star forming gas cloud.
Star formation in the nuclear region can occur in several dense star-forming regions (see, e.g., Figure~\ref{fig:young_star_dist}).
However, for feedback purposes, massive young stars quickly separate from their birth clouds and the {\it entire} nuclear region therefore acts as a single short dynamical time region. 

Observed nuclear stellar disks show indications of oscillatory star formation and feedback cycles.
\citet{Davies2007} found that a sample of Seyfert galaxies contained nuclear star clusters with spatial extent $\ltsim$ 50 pc.
Interestingly, although these nuclear star clusters could be identified to have characteristic ages of $\sim$10-300 Myr, 
there was little or no evidence of ongoing star formation in these same systems (as inferred through low values for the equivalent widths of Br$\gamma$ emission).
We caution, however, that the modest gas surface densities ($\Sigma_{{\rm gas}}\ltsim 300 M_\odot$/pc${}^2$) explored in this paper are not directly comparable to the population ($\Sigma_{{\rm gas}}\gtrsim 1000 M_\odot$/pc${}^2$) studied in~\citet{Davies2007}.

If there is a link between aging stellar populations and the fueling of AGN activity, as has been implied from the time offset between nuclear star formation and Seyfert activity, then the bursty nature of star formation found in our model would have implications for AGN fueling.  
Our models included a central supermassive black hole -- which is dynamically important for the innermost galactic regions studied in this paper -- as well as stellar mass loss.
However, we elected to not include black hole growth and associated black hole feedback in the present study as both processes add significant complicating factors to the interpretation of the results.
In the future, we intend to run additional simulations that consider the relationship between nuclear star formation and AGN activity~\citep[for preliminary results on black hole growth and feedback in simulations focusing on the inner $\sim$100 pc regions of gas-rich galaxies, see][]{Hopkins2015}.

A particularly  interesting application of full galaxy simulations that will include both explicit feedback from stars and from massive black holes in a resolved ISM will be to study the generation and effects of galaxy-scale outflows driven by luminous AGN. 
Such outflows have now been detected in luminous quasars over a wide range of redshift and in both atomic and molecular gas~\citep[e.g.,][]{Rupke2011, Greene2012, Cicone2014, Harrison2014}.
These appear energetic enough to potentially strongly affect galaxy scale star formation. 
While idealized analytic and numerical calculations~\citep[e.g.,][]{FaucherGiguere2012, Zubovas2012, Costa2014, Bourne2014} have begun to address the physics of such outflows, fully dynamical simulations of galaxies including both stellar and black hole feedback will likely be needed to robustly predicted how AGN-driven galactic winds are mass loaded and how they affect the host galaxy.

\section{Conclusions}
\label{sec:Conclusions}
We have studied the impact of feedback from young stars on the star formation rates and gas content of galactic nuclei.  
We outlined basic analytic arguments that modify existing feedback equilibrium star formation models~\citep[e.g.,][]{Thompson2005, Ostriker2011, FaucherGiguere2013} to apply to short dynamical time regions.
We ran and explored numerical simulations that included realistic, time resolved stellar feedback prescriptions to directly model the evolution of several galaxy models.

Our primary conclusions are as follows:
\begin{itemize}
\item The nuclear regions ($r\ltsim100$ pc) of galaxies are characterized by dynamical times of order $\sim$1 Myr.  
Feedback from young stars evolves over timescales of tens of Myr (Figure~\ref{fig:sb99}).  
This timescale mismatch means that feedback from young stars is not immediately responsive to changes in the gas state, and thus there is no {\it stable equilibrium star formation rate}.
\item Instead, gas will continue to form stars until a critical mass of young stars is formed (eqn.~\ref{eqn:NewFeedbackEquilibrium}), at which point the feedback from young stars leads to an unstable blowout from the nucleus.
After the blowout new gas flows into the nucleus from larger radii (Figure~\ref{fig:gas_surface_density_2}).
As a consequence, the gas content, star formation rate, and mass of young stars all oscillate in time when measured over $\sim 100$ pc scales; they are significantly more stable when measured over larger $\gtrsim 1$ kpc scales (Figures~\ref{fig:gas_nuclear_evolution} and~\ref{fig:gas_nuclear_evolution2}).
\item The simulated KS relation shows very different properties on $100$ pc and $1$ kpc scales (see Figure~\ref{fig:ks_relation1}).  
The long dynamical time on large scales means the system can reach equilibrium. 
This leads to a more stable star formation rate and gas content with time -- hence a tighter KS relation.
In the galactic nucleus, by contrast, the KS relation is much more variable and has larger scatter.
\end{itemize}

In the present study the dynamical effect of the central supermassive black hole was included but AGN feedback was not.
In this paper we have shown that stellar feedback {\it alone} is capable of driving episodic star formation and outflows from galactic nuclei.
However, considering both stellar and AGN feedback in concert will be important for understanding the relative importance of different feedback channels on galactic nuclei.

\section*{Acknowledgements} 
PT acknowledges helpful discussions with Sara Ellison, Nick McConnell, and Sarah Wellons.
PT acknowledges support from NASA ATP Grant NNX14AH35G (PI Hopkins) and support through an MIT RSC award.
Support for PFH was provided by an Alfred P. Sloan Research Fellowship, NASA ATP Grant NNX14AH35G, NSF Collaborative Research Grant \#1411920, and CAREER grant \#1455342. 
CAFG was supported by NSF through grants AST-1412836 and AST-1517491, by NASA through grant NNX15AB22G, and by STScI through grant HST-AR-14293.001-A. 
MV acknowledges support through an MIT RSC award.
DK was supported by NSF grant AST-1412153.
EQ was supported in part by NASA ATP grant 12-APT12-0183, a Simons Investigator award from the Simons Foundation, and the David and Lucile Packard Foundation.
The simulations reported in this paper were run and processed on the ``Quest" computer cluster at Northwestern University,  the Caltech compute cluster ``Zwicky" (NSF MRI award \#PHY-0960291), the joint partition of the MIT-Harvard computing cluster ``Odyssey" supported by MKI and FAS, and allocation TG-AST130039 and TG-AST150059 granted by the Extreme Science and Engineering Discovery Environment (XSEDE) supported by the NSF.


\begin{thebibliography}{58}
\expandafter\ifx\csname natexlab\endcsname\relax\def\natexlab#1{#1}\fi

\bibitem[{{Bigiel} {et~al.}(2008){Bigiel}, {Leroy}, {Walter}, {Brinks}, {de
  Blok}, {Madore}, \& {Thornley}}]{Bigiel2008}
{Bigiel}, F., {Leroy}, A., {Walter}, F., {et~al.} 2008, \aj, 136, 2846

\bibitem[{{Booth} {et~al.}(2013){Booth}, {Agertz}, {Kravtsov}, \&
  {Gnedin}}]{Booth2013}
{Booth}, C.~M., {Agertz}, O., {Kravtsov}, A.~V., \& {Gnedin}, N.~Y. 2013,
  \apjl, 777, L16

\bibitem[{{Bouch{\'e}} {et~al.}(2007){Bouch{\'e}}, {Cresci}, {Davies},
  {Eisenhauer}, {F{\"o}rster Schreiber}, {Genzel}, {Gillessen}, {Lehnert},
  {Lutz}, {Nesvadba}, {Shapiro}, {Sternberg}, {Tacconi}, {Verma}, {Cimatti},
  {Daddi}, {Renzini}, {Erb}, {Shapley}, \& {Steidel}}]{Bouche2007}
{Bouch{\'e}}, N., {Cresci}, G., {Davies}, R., {et~al.} 2007, \apj, 671, 303

\bibitem[{{Bourne} {et~al.}(2014){Bourne}, {Nayakshin}, \&
  {Hobbs}}]{Bourne2014}
{Bourne}, M.~A., {Nayakshin}, S., \& {Hobbs}, A. 2014, \mnras, 441, 3055

\bibitem[{{Cicone} {et~al.}(2014){Cicone}, {Maiolino}, {Sturm},
  {Graci{\'a}-Carpio}, {Feruglio}, {Neri}, {Aalto}, {Davies}, {Fiore},
  {Fischer}, {Garc{\'{\i}}a-Burillo}, {Gonz{\'a}lez-Alfonso},
  {Hailey-Dunsheath}, {Piconcelli}, \& {Veilleux}}]{Cicone2014}
{Cicone}, C., {Maiolino}, R., {Sturm}, E., {et~al.} 2014, \aap, 562, A21

\bibitem[{{Cioffi} {et~al.}(1988){Cioffi}, {McKee}, \&
  {Bertschinger}}]{Cioffi1988}
{Cioffi}, D.~F., {McKee}, C.~F., \& {Bertschinger}, E. 1988, \apj, 334, 252

\bibitem[{{Costa} {et~al.}(2014){Costa}, {Sijacki}, \& {Haehnelt}}]{Costa2014}
{Costa}, T., {Sijacki}, D., \& {Haehnelt}, M.~G. 2014, \mnras, 444, 2355

\bibitem[{{Davies} {et~al.}(2007){Davies}, {M{\"u}ller S{\'a}nchez}, {Genzel},
  {Tacconi}, {Hicks}, {Friedrich}, \& {Sternberg}}]{Davies2007}
{Davies}, R.~I., {M{\"u}ller S{\'a}nchez}, F., {Genzel}, R., {et~al.} 2007,
  \apj, 671, 1388

\bibitem[{{Dolag} {et~al.}(1999){Dolag}, {Bartelmann}, \& {Lesch}}]{Dolag1999}
{Dolag}, K., {Bartelmann}, M., \& {Lesch}, H. 1999, \aap, 348, 351

\bibitem[{{Fall} {et~al.}(2010){Fall}, {Krumholz}, \& {Matzner}}]{Fall2010}
{Fall}, S.~M., {Krumholz}, M.~R., \& {Matzner}, C.~D. 2010, \apjl, 710, L142

\bibitem[{{Faucher-Gigu{\`e}re} {et~al.}(2015){Faucher-Gigu{\`e}re}, {Hopkins},
  {Kere{\v s}}, {Muratov}, {Quataert}, \& {Murray}}]{FaucherGiguere2015}
{Faucher-Gigu{\`e}re}, C.-A., {Hopkins}, P.~F., {Kere{\v s}}, D., {et~al.}
  2015, \mnras, 449, 987

\bibitem[{{Faucher-Gigu{\`e}re} \& {Quataert}(2012)}]{FaucherGiguere2012}
{Faucher-Gigu{\`e}re}, C.-A., \& {Quataert}, E. 2012, \mnras, 425, 605

\bibitem[{{Faucher-Gigu{\`e}re} {et~al.}(2013){Faucher-Gigu{\`e}re},
  {Quataert}, \& {Hopkins}}]{FaucherGiguere2013}
{Faucher-Gigu{\`e}re}, C.-A., {Quataert}, E., \& {Hopkins}, P.~F. 2013, \mnras,
  433, 1970

\bibitem[{{Federrath} \& {Klessen}(2012)}]{Federrath2012}
{Federrath}, C., \& {Klessen}, R.~S. 2012, \apj, 761, 156

\bibitem[{{Gatto} {et~al.}(2015){Gatto}, {Walch}, {Low}, {Naab}, {Girichidis},
  {Glover}, {W{\"u}nsch}, {Klessen}, {Clark}, {Baczynski}, {Peters},
  {Ostriker}, {Ib{\'a}{\~n}ez-Mej{\'{\i}}a}, \& {Haid}}]{Gatto2015}
{Gatto}, A., {Walch}, S., {Low}, M.-M.~M., {et~al.} 2015, \mnras, 449, 1057

\bibitem[{{Genzel} {et~al.}(2010){Genzel}, {Tacconi}, {Gracia-Carpio},
  {Sternberg}, {Cooper}, {Shapiro}, {Bolatto}, {Bouch{\'e}}, {Bournaud},
  {Burkert}, {Combes}, {Comerford}, {Cox}, {Davis}, {Schreiber},
  {Garcia-Burillo}, {Lutz}, {Naab}, {Neri}, {Omont}, {Shapley}, \&
  {Weiner}}]{Genzel2010}
{Genzel}, R., {Tacconi}, L.~J., {Gracia-Carpio}, J., {et~al.} 2010, \mnras,
  407, 2091

\bibitem[{{Greene} {et~al.}(2012){Greene}, {Zakamska}, \& {Smith}}]{Greene2012}
{Greene}, J.~E., {Zakamska}, N.~L., \& {Smith}, P.~S. 2012, \apj, 746, 86

\bibitem[{{Harrison} {et~al.}(2014){Harrison}, {Alexander}, {Mullaney}, \&
  {Swinbank}}]{Harrison2014}
{Harrison}, C.~M., {Alexander}, D.~M., {Mullaney}, J.~R., \& {Swinbank}, A.~M.
  2014, \mnras, 441, 3306

\bibitem[{{Hernquist}(1990)}]{H90}
{Hernquist}, L. 1990, \apj, 356, 359

\bibitem[{{Hopkins}(2015)}]{GIZMO}
{Hopkins}, P.~F. 2015, \mnras, 450, 53

\bibitem[{{Hopkins} {et~al.}(2014){Hopkins}, {Kere{\v s}}, {O{\~n}orbe},
  {Faucher-Gigu{\`e}re}, {Quataert}, {Murray}, \& {Bullock}}]{Hopkins2014}
{Hopkins}, P.~F., {Kere{\v s}}, D., {O{\~n}orbe}, J., {et~al.} 2014, \mnras,
  445, 581

\bibitem[{{Hopkins} {et~al.}(2010){Hopkins}, {Murray}, {Quataert}, \&
  {Thompson}}]{Hopkins2010b}
{Hopkins}, P.~F., {Murray}, N., {Quataert}, E., \& {Thompson}, T.~A. 2010,
  \mnras, 401, L19

\bibitem[{{Hopkins} \& {Quataert}(2011)}]{Hopkins2011c}
{Hopkins}, P.~F., \& {Quataert}, E. 2011, \mnras, 415, 1027

\bibitem[{{Hopkins} {et~al.}(2011{\natexlab{a}}){Hopkins}, {Quataert}, \&
  {Murray}}]{Hopkins2011a}
{Hopkins}, P.~F., {Quataert}, E., \& {Murray}, N. 2011{\natexlab{a}}, \mnras,
  417, 950

\bibitem[{{Hopkins} {et~al.}(2011{\natexlab{b}}){Hopkins}, {Quataert}, \&
  {Murray}}]{HopkinsKSlaw}
---. 2011{\natexlab{b}}, \mnras, 417, 950

\bibitem[{{Hopkins} {et~al.}(2015){Hopkins}, {Torrey}, {Faucher-Giguere},
  {Quataert}, \& {Murray}}]{Hopkins2015}
{Hopkins}, P.~F., {Torrey}, P., {Faucher-Giguere}, C.-A., {Quataert}, E., \&
  {Murray}, N. 2015, ArXiv e-prints

\bibitem[{{Jubelgas} {et~al.}(2008){Jubelgas}, {Springel}, {En{\ss}lin}, \&
  {Pfrommer}}]{Jubelgas2008}
{Jubelgas}, M., {Springel}, V., {En{\ss}lin}, T., \& {Pfrommer}, C. 2008, \aap,
  481, 33

\bibitem[{{Kennicutt} \& {Evans}(2012)}]{Kennicutt2012}
{Kennicutt}, R.~C., \& {Evans}, N.~J. 2012, \araa, 50, 531

\bibitem[{{Kennicutt}(1998)}]{Kennicutt}
{Kennicutt}, Jr., R.~C. 1998, \apj, 498, 541

\bibitem[{{Kim} \& {Ostriker}(2015)}]{Kim2015}
{Kim}, C.-G., \& {Ostriker}, E.~C. 2015, \apj, 802, 99

\bibitem[{{Krumholz} \& {Gnedin}(2011)}]{Krumholz2011}
{Krumholz}, M.~R., \& {Gnedin}, N.~Y. 2011, \apj, 729, 36

\bibitem[{{Krumholz} {et~al.}(2006){Krumholz}, {Matzner}, \&
  {McKee}}]{Krumholz2006}
{Krumholz}, M.~R., {Matzner}, C.~D., \& {McKee}, C.~F. 2006, \apj, 653, 361

\bibitem[{{Krumholz} \& {McKee}(2005)}]{Krumholz2005}
{Krumholz}, M.~R., \& {McKee}, C.~F. 2005, \apj, 630, 250

\bibitem[{{Leitherer} {et~al.}(2010){Leitherer}, {Ortiz Ot{\'a}lvaro},
  {Bresolin}, {Kudritzki}, {Lo Faro}, {Pauldrach}, {Pettini}, \&
  {Rix}}]{SB99_3}
{Leitherer}, C., {Ortiz Ot{\'a}lvaro}, P.~A., {Bresolin}, F., {et~al.} 2010,
  \apjs, 189, 309

\bibitem[{{Leroy} {et~al.}(2008){Leroy}, {Walter}, {Brinks}, {Bigiel}, {de
  Blok}, {Madore}, \& {Thornley}}]{Leroy2008}
{Leroy}, A.~K., {Walter}, F., {Brinks}, E., {et~al.} 2008, \aj, 136, 2782

\bibitem[{{Marinacci} {et~al.}(2015){Marinacci}, {Vogelsberger}, {Mocz}, \&
  {Pakmor}}]{Marinacci2015}
{Marinacci}, F., {Vogelsberger}, M., {Mocz}, P., \& {Pakmor}, R. 2015, ArXiv
  e-prints

\bibitem[{{Martizzi} {et~al.}(2015){Martizzi}, {Faucher-Gigu{\`e}re}, \&
  {Quataert}}]{Martizzi2015}
{Martizzi}, D., {Faucher-Gigu{\`e}re}, C.-A., \& {Quataert}, E. 2015, \mnras,
  450, 504

\bibitem[{{Mo} {et~al.}(1998){Mo}, {Mao}, \& {White}}]{MMWdisk}
{Mo}, H.~J., {Mao}, S., \& {White}, S.~D.~M. 1998, \mnras, 295, 319

\bibitem[{{Muratov} {et~al.}(2015){Muratov}, {Keres}, {Faucher-Giguere},
  {Hopkins}, {Quataert}, \& {Murray}}]{Muratov2015}
{Muratov}, A.~L., {Keres}, D., {Faucher-Giguere}, C.-A., {et~al.} 2015, ArXiv
  e-prints

\bibitem[{{Murray} {et~al.}(2010){Murray}, {Quataert}, \&
  {Thompson}}]{Murray2010}
{Murray}, N., {Quataert}, E., \& {Thompson}, T.~A. 2010, \apj, 709, 191

\bibitem[{{Narayanan} {et~al.}(2012){Narayanan}, {Krumholz}, {Ostriker}, \&
  {Hernquist}}]{Narayanan2012}
{Narayanan}, D., {Krumholz}, M.~R., {Ostriker}, E.~C., \& {Hernquist}, L. 2012,
  \mnras, 421, 3127

\bibitem[{{Navarro} {et~al.}(1997){Navarro}, {Frenk}, \& {White}}]{NFW}
{Navarro}, J.~F., {Frenk}, C.~S., \& {White}, S.~D.~M. 1997, \apj, 490, 493

\bibitem[{{Ostriker} \& {Shetty}(2011)}]{Ostriker2011}
{Ostriker}, E.~C., \& {Shetty}, R. 2011, \apj, 731, 41

\bibitem[{{Pakmor} \& {Springel}(2013)}]{Pakmor2013}
{Pakmor}, R., \& {Springel}, V. 2013, \mnras, 432, 176

\bibitem[{{Rupke} \& {Veilleux}(2011)}]{Rupke2011}
{Rupke}, D.~S.~N., \& {Veilleux}, S. 2011, \apjl, 729, L27

\bibitem[{{Salem} {et~al.}(2014){Salem}, {Bryan}, \& {Hummels}}]{Salem2014}
{Salem}, M., {Bryan}, G.~L., \& {Hummels}, C. 2014, \apjl, 797, L18

\bibitem[{{Sales} {et~al.}(2014){Sales}, {Marinacci}, {Springel}, \&
  {Petkova}}]{Sales2014}
{Sales}, L.~V., {Marinacci}, F., {Springel}, V., \& {Petkova}, M. 2014, \mnras,
  439, 2990

\bibitem[{{Schmidt}(1959)}]{Schmidt}
{Schmidt}, M. 1959, \apj, 129, 243

\bibitem[{{Silk}(1997)}]{Silk1997}
{Silk}, J. 1997, \apj, 481, 703

\bibitem[{{Springel}(2005)}]{GADGET}
{Springel}, V. 2005, \mnras, 364, 1105

\bibitem[{{Springel} \& {White}(1999)}]{SpringelWhiteDisk}
{Springel}, V., \& {White}, S.~D.~M. 1999, \mnras, 307, 162

\bibitem[{{Thompson} {et~al.}(2005){Thompson}, {Quataert}, \&
  {Murray}}]{Thompson2005}
{Thompson}, T.~A., {Quataert}, E., \& {Murray}, N. 2005, \apj, 630, 167

\bibitem[{{Uhlig} {et~al.}(2012){Uhlig}, {Pfrommer}, {Sharma}, {Nath},
  {En{\ss}lin}, \& {Springel}}]{Uhlig2012}
{Uhlig}, M., {Pfrommer}, C., {Sharma}, M., {et~al.} 2012, \mnras, 423, 2374

\bibitem[{{Walch} {et~al.}(2012){Walch}, {Whitworth}, {Bisbas}, {W{\"u}nsch},
  \& {Hubber}}]{Walch2012}
{Walch}, S.~K., {Whitworth}, A.~P., {Bisbas}, T., {W{\"u}nsch}, R., \&
  {Hubber}, D. 2012, \mnras, 427, 625

\bibitem[{{Wang} \& {Abel}(2009)}]{Wang2009}
{Wang}, P., \& {Abel}, T. 2009, \apj, 696, 96

\bibitem[{{Whitworth}(1979)}]{Whitworth1979}
{Whitworth}, A. 1979, \mnras, 186, 59

\bibitem[{{Zubovas} \& {King}(2012)}]{Zubovas2012}
{Zubovas}, K., \& {King}, A. 2012, \apjl, 745, L34

\bibitem[{{Zuckerman} \& {Evans}(1974)}]{Zuckerman1974}
{Zuckerman}, B., \& {Evans}, II, N.~J. 1974, \apjl, 192, L149

\end{thebibliography}

\end{document}